\documentclass[a4paper,fleqn,usenatbib]{mnras}
\usepackage[T1]{fontenc}
\usepackage{ae,aecompl}
\usepackage{amsmath}
\usepackage{graphicx}
\usepackage{mathrsfs}
\usepackage{color}
\usepackage{epstopdf}
\usepackage{hyperref}
\usepackage{hypcap}
\DeclareMathOperator{\arctanh}{arctanh}
\DeclareMathOperator{\kpc}{kpc}

\title[A new method to break the mass sheet degeneracy using aperture moments]{A new method to break the mass sheet degeneracy using aperture moments}
\author[Rexroth, Natarajan, \& Kneib]{Markus Rexroth$^{1}$, Priyamvada Natarajan$^{2}$, \& Jean-Paul Kneib$^{1,3}$\\
\\
$^{1}$ Laboratoire d'Astrophysique, \'{E}cole Polytechnique F\'{e}d\'{e}rale de Lausanne (EPFL), Observatoire de Sauverny, CH-1290 Versoix, Switzerland\\
$^{2}$ Department of Astronomy, Yale University, 260 Whitney Avenue, New Haven, CT 06511, USA \\
$^{3}$ Aix Marseille Universit\'{e}, CNRS, LAM (Laboratoire d'Astrophysique de Marseille) UMR 7326, 13388, Marseille, France}

\date{Accepted XXX. Received YYY; in original form ZZZ}
\pubyear{2016}

\begin{document}
\label{firstpage}
\pagerange{\pageref{firstpage}--\pageref{lastpage}}
\maketitle

\begin{abstract}
Mass determinations from gravitational lensing shear and the higher order estimator flexion are both subject to the mass sheet degeneracy. Mass sheet degeneracy refers to a transformation that leaves the reduced shear and flexion invariant. In general, this transformation can be approximated by the addition of a constant surface mass density sheet. We propose a new technique to break the mass sheet degeneracy. The method uses mass moments of the shear or flexion fields in combination with convergence information derived from number counts which exploit the magnification bias. The difference between the measured mass moments provides an estimator for the magnitude of the additive constant that is the mass-sheet. For demonstrating this, we derive relations that hold true in general for \textit{n}-th order moments and show how they can be employed effectively to break the degeneracy. We investigate the detectability of this degeneracy parameter from our method and find that the degeneracy parameter can be feasibly determined from stacked galaxy-galaxy lensing data and cluster lensing data. Furthermore, we compare the signal-to-noise ratios of convergence information from number counts with shear and flexion for SIS and NFW models. We find that the combination of shear and flexion performs best on galaxy and cluster scales and the convergence information can therefore be used to break the mass sheet degeneracy without quality loss in the mass reconstruction. In summary, there is power in the combination of shear, flexion, convergence and their higher order moments. With the anticipated wealth of lensing data from upcoming and future satellite missions - EUCLID and WFIRST - this technique will be feasible.
\end{abstract}


\section{Introduction}

Many mass reconstructions were successfully obtained utilizing gravitational lensing shear \citep[e.g.,][]{Hoekstra2004,Natarajan1998} and the feasibility of lens models from flexion has been demonstrated \citep{Leonard2007,Okura2008}. Several mass reconstruction methods have been developed \citep[e.g.,][]{Schneider2000,Bartelmann1996a,Kaiser1995a, Kaiser1993}, see \citet{Kneib2011} for a review of galaxy cluster lens reconstructions. However, the masses derived from the shear and flexion fields are only determined up to a constant due to the mass sheet degeneracy \citep[e.g.,][]{Schneider1995,Schneider2008}. Several techniques have been proposed to lift the degeneracy. \citet{Broadhurst1995} suggested using magnification information by comparing lensed and unlensed background source counts to reconstruct the non-degenerate mass sheet. This effect was subsequently observed and applied \citep[e.g.][]{Fort1997,Taylor1998,Dye2002}. \citet{Bartelmann1995} proposed the lens parallax method. This method compares the mean sizes of lensed faint blue galaxies with those of unlensed sources in an empty control field. While gravitational lensing magnifies the area, it preserves the surface brightness, and thus the magnification can be inferred by comparing the mean sizes of lensed and unlensed galaxies with the same surface brightness. The magnification information can then be combined with the shear measurement to break the degeneracy. However, it is hard to estimate the surface brightness from seeing-convolved images and thus the application of this technique is challenging \citep{Schneider2006}. In addition, the degeneracy can also be lifted by fitting the gravitational lens potential to shear and magnification data simultaneously. Several fitting and Bayesian methods have been proposed for this purpose \citep[see e.g.][]{Umetsu2011,Schneider2000,Bartelmann1996a}. \citet{Bradavc2004} showed that it is possible to reconstruct non-degenerate mass distributions if the individual redshifts of sources at different distances are combined with the shear field. However, this technique can only be used for critical lenses. A different method is needed for sub-critical lenses and lenses where a sufficient number of redshifts is not available. In this paper, we present, develop, and explicate the technical framework that can be used to derive the mass sheet degeneracy using combinations of shape and convergence estimator moments.  \\

We propose a new method that can break the mass-sheet degeneracy by a simple comparison of aperture masses. We show that taking the difference in mass estimates from convergence information, derived for example from source counts, and shear or flexion data allows the determination of the value of the mass sheet, thus enabling the derivation of calibrated masses. We demonstrate its feasibility for both stacked galaxy-galaxy lensing and cluster lensing. Furthermore, we show that for two widely used mass models, the signal-to-noise of shear and flexion in the weak lensing regime is always superior to the number counts, both on galaxy and cluster scales. Therefore we can use the convergence information to break the mass sheet without sacrificing the quality of the lens reconstruction. \\

We develop the theoretical framework of this method for both shear and flexion, the higher order lensing effect which describes the arciness of the lensed sources. Flexion was introduced into the weak lensing framework, because it provides valuable additional information and is highly sensitive to substructure \citep{Bacon2006,Irwin2006,Goldberg2005,Goldberg2002}. Studies have showed that the addition of flexion can improve mass reconstructions significantly \citep[e.g.,][]{Leonard2007,Okura2007}. The high quality data from the \textit{Hubble Frontier Fields} initiative permit flexion measurements with increased accuracy \citep{Rexroth2015}. \\

The paper is organized as follows. In section 2, we give a brief description of the weak lensing formalism and the mass-sheet degeneracy. Section 3 presents the equivalent moments and how their transformations can break the degeneracy. We demonstrate this method for a singular isothermal sphere (SIS) halo model in section 4. In section 5 we investigate the detectability of the mass sheet parameter without making assumptions on the halo model. Section 6 shows that the combination of shear and flexion has a higher signal-to-noise ratio than number counts for SIS and Navarro-Frenk-White (NFW) halos. We conclude in section 7.

\section{Weak lensing formalism and the mass-sheet degeneracy}

Weak lensing of background sources by massive foreground objects results in the deformation of their shapes. The strength of these distortions is directly related to the surface mass density of the lens. Let us consider a lens with projected surface mass density $\Sigma(\boldsymbol{\theta})$, where the angular coordinate $\boldsymbol{\theta}$ denotes the position in the lens plane. The convergence of the lens $\kappa$ is defined as 
\begin{eqnarray}
\kappa(\boldsymbol{\theta}) = \frac{\Sigma(\boldsymbol{\theta})}{\Sigma_{\rm crit}},
\end{eqnarray} 
where the critical surface mass density, defined as
\begin{eqnarray}
\Sigma_{\rm crit}\,=\,{\frac{c^2}{4 \pi G}}\, \Big(\frac{D_{\rm os}}{D_{\rm ls}\,D_{\rm ol}}\Big),
\end{eqnarray}
depends on the angular diameter distances from the lens to the source $D_{\rm ls}$, observer to source
$D_{\rm os}$, and observer to lens $D_{\rm ol}$. The lensed and unlensed coordinates for the distorted background 
sources are related by the mapping
\begin{eqnarray}
A_{ij}(\boldsymbol{\theta}) \,=\, \frac{\partial \theta^{'}_{i}}{\partial \theta_{j}}, \\
A(\boldsymbol{\theta}) \,=\, \bigg(
\begin{matrix}
1 - \kappa - \gamma_{1} & - \gamma_{2} \\
- \gamma_{2} & 1 - \kappa + \gamma_{1} 
\end{matrix} \bigg),
\end{eqnarray}
where $\boldsymbol{\theta^{'}}$ are the unlensed coordinates and $\gamma$ is the shear. The magnification of the sources is 
\begin{align}
\mu = \frac{1}{(1-\kappa)^2-\gamma^2}.
\end{align}
Furthermore, we can express the convergence and shear using the deflection potential 
\begin{align}
\psi(\boldsymbol{\theta}) = \frac{1}{\pi} \int d^2\beta~\kappa(\boldsymbol{\beta}) \ln|\boldsymbol{\theta} - \boldsymbol{\beta}|
\end{align}
as
\begin{align}
\kappa \,=\, \frac{1}{2} (\psi_{,11} + \psi_{,22}),\,\, \gamma \,=\, \bigg(
\begin{matrix}
\frac{1}{2} (\psi_{,11} - \psi_{,22}) \\
\psi_{,12}
\end{matrix} \bigg).
\end{align}
The commas denote partial $\theta$ derivatives. The reduced shear is defined as 
\begin{align}
g \,=\, \frac{\gamma}{1 - \kappa}.
\end{align} 
If shear and convergence are constant over a source image, the transformation is given by a simple matrix multiplication:
\begin{eqnarray}
\theta^{'}_{i} \,=\, A_{ij}\, \theta_{j}.
\end{eqnarray}
However, if the shear varies over the image, we have to expand to include flexion terms:
\begin{eqnarray}
\theta^{'}_{i} \,\simeq\, A_{ij}\, \theta_{j} + \frac{1}{2} D_{ijk}\, \theta_{j}\, \theta_{k},
\end{eqnarray}
where $D_{ijk} \,=\, A_{ij,k}$. The first and second flexion are given in terms of derivatives of the two shear components \citep[as per the notation in][]{Bacon2006}
\begin{align}
\mathcal{F} \,=\, \bigg(
\begin{matrix}
\gamma_{1,1} + \gamma_{2,2} \\
\gamma_{2,1} - \gamma_{1,2}
\end{matrix} \bigg),\,\,
\mathcal{G} \,=\, \bigg(
\begin{matrix}
\gamma_{1,1} - \gamma_{2,2} \\
\gamma_{2,1} + \gamma_{1,2} 
\end{matrix} \bigg). \label{Def_flexion_shear_deriv}
\end{align}

The mass sheet degeneracy arises due to the fact that the observed quantities are not the shear and flexions themselves, but the reduced shear and the reduced first and second flexion. The latter two can be compactly written if we use a complex notation, i.e. $\mathcal{F} \,=\, \mathcal{F}_{1} + i\, \mathcal{F}_{2}$, 
\begin{align}
 G_{1} \,=\, \frac{\mathcal{F} + g\, \mathcal{F}^{*}}{1 - \kappa},\,\, G_{3} \,=\, \frac{\mathcal{G} + g\, \mathcal{F}}{1 - \kappa},
\end{align}
where the asterisk denotes complex conjugation \citep{Schneider2008}. Alternatively, the reduced flexions can be defined as \citep{Okura2008}
\begin{align}
F\,=\,\frac{\mathcal{F}}{1-\kappa},\,\,
G\,=\,\frac{\mathcal{G}}{1-\kappa},
\end{align} 
and we will use this definition in this paper. If the deflection potential is transformed for any constant $\lambda$ as
\begin{eqnarray}
\psi(\boldsymbol{\theta}) \to \psi'(\boldsymbol{\theta}) \,=\, \frac{1 - \lambda}{2} \boldsymbol{\theta}^{2} + \lambda\, \psi(\boldsymbol{\theta}),
\end{eqnarray}
we have
\begin{align}
\kappa \rightarrow \kappa'\,=\,\lambda\,\kappa\,+\,(1 - \lambda),\,\,
\gamma\,\rightarrow \gamma'\,=\,\lambda\,\gamma, \label{Intro_transforms} \\
\mathcal{F} \rightarrow \mathcal{F}' \,=\, \lambda\, \mathcal{F},\,\, \mathcal{G} \rightarrow \mathcal{G}' \,=\, \lambda\, \mathcal{G},\,\,
\mu \rightarrow \mu'\,=\,\frac{\mu}{\lambda^2}, \nonumber 
\end{align}
but the reduced shear and reduced flexions are invariant under such a family of transformations,
\begin{align}
g \rightarrow g'\,=\,g,\,\, F \rightarrow F'\,=\,F,\,\, G \rightarrow G'\,=\,G.\label{Intro_reduced_transforms}
\end{align} 
Therefore two different surface mass distributions that differ by $\lambda$ cannot be distinguished by measuring only the image distortions. Note that in the limit of $\lambda$ almost equal to unity, equation~(\ref{Intro_transforms}) amounts to adding a constant mass sheet $\kappa_0$ to $\kappa$.\\

In principle, the mass-sheet degeneracy could be lifted by requiring that $\kappa \,=\, 0$ at the boundary of the image. However, even for wide-field cameras this leads to a substantial underestimate of the cluster mass as cluster density profiles appear to be falling smoothly even out to large radius. For a cluster with virial mass $M_{\rm vir} \,=\, 10^{15}\, M_{\odot}$ at redshift $z \,=\, 0.2$, we expect from N-body simulations that $\kappa \,\approx\, 0.005$ at $15 \arcmin$ from the cluster center and with a $30 \arcmin \times 30\arcmin$ camera field of view, we would underestimate the virial mass by $\sim 20 \%$ if we simply set $\kappa$ to zero at the boundary \citep{Bradavc2004}. Therefore more sophisticated techniques are required.

\section{Proposed method: Moments of the convergence, shear and flexion fields}

Our goal is to derive a method that is independent of spectroscopic redshifts and broadly applicable to both critical and sub-critical lenses. To develop such a method, we first investigate $n$-th order moments of the 
shear, convergence, and flexions and study their transformation properties with a view to understanding the explicit dependence on $\lambda$. \citet{Schneider1997} (hereafter called SB97) show the equivalence of shear and convergence mass moments $M^{(n)}$ and multipole moments $Q^{(n)}$. We extend these equivalences to mass and multipole moments using flexion. Subsequently, we show that the mass-sheet degeneracy transformations destroy these equivalences, as they give rise to different surface terms. These permit us to calculate the mass-sheet degeneracy parameter $\lambda$.\\

Our method comes in two variants. One uses the reconstructed, unreduced shear or flexion fields. We can obtain these by multiplying the reduced quantities with $1-\kappa_{\rm rec}$, where $\kappa_{\rm rec}$ is the convergence of the reconstructed mass-sheet. Many techniques for obtaining $\kappa_{\rm rec}$ from shear or flexion data have been successfully developed and \citet{Er2010} present a method that reconstructs the mass-sheet by combining the two. The second variant uses the reduced shear and flexion fields, thus it does not require the reconstructed convergence.\\

\subsection{Mass and multipole moments of shear, flexion, and convergence}
The detailed derivation of the different moments and the proof of their equivalences is shown in appendix~\ref{Appendix:Non-reduced moments of the full field of view}. We do not explicitly denote $(x_0,\varphi_0)$ or $(x,\varphi)$ dependencies of $\kappa,\kappa_{rec},\gamma,\mathcal{F}$, and $\mathcal{G}$ to keep the notation compact. The resulting equivalent moments are:
\begin{align}
M^{(n)}_{\kappa}\,=\,\int_0^{\infty}\, dx\,\, x^{n+1}\, w(x)\, \int_0^{2\pi}\, d\varphi\,\, \kappa,\\
M^{(n)}_{\mathcal{F}}\,=\, \int_0^{\infty}\, dx\,\, x\,W(x)\,\int_0^{2\pi}\,d\varphi\,\, (1-\kappa_{\rm rec}) F_{t},\\
M^{(n)}_{\gamma} \,=\, \int\limits_{0}^{\infty}\, dx\,\, \big[ 2 W(x) - x^{n+1}\, w(x) \big] \int\limits_{0}^{2 \pi}\, d\varphi \nonumber \\
(1-\kappa_{\rm rec}) g_{t},\\
M^{(n)}_{\mathcal{F},\mathcal{G}} \,=\, -\int_0^{\infty}\, dx\,\, \big[V(x)\, - \frac{1}{2} x\, W(x) \big]\,  \int_0^{2\pi}\,d\varphi \nonumber \\  \big[ (1-\kappa_{\rm rec}) F_{t} + (1-\kappa_{\rm rec}) G_{t} \big].
\end{align}
The formula for $M^{(n)}_{\mathcal{F}}$ in the case that $n=0$ was also derived in \citet{Leonard2009}, but with stronger assumptions on the weighting function. We used the definitions
\begin{align}
x\, W(x) \,=\, \int\limits_{0}^{x} \, dy\,\, y^{n+1}\, w(y),\\
V(x) \,=\, \int\limits_{0}^{x}\, dy\,\, W(y),
\end{align}
and the reduced tangential and radial shear and flexions,
\begin{align} 
g_{t} \,=\, -\big[g_{1}(x,\varphi)\, \cos(2\varphi) + g_{2}(x,\varphi)\, \sin(2\varphi) \big], \\
g_{r} \,=\, -\big[g_{2}(x,\varphi)\, \cos(2\varphi) - g_{1}(x,\varphi)\, \sin(2\varphi) \big],\\
F_t\,=\,-\big[F_1(x,\varphi) \cos(\varphi) + F_2(x,\varphi) \sin(\varphi) \big],\\
F_{r} \,=\, -\big[ F_{2}(x,\varphi)\, \cos(\varphi) - F_{1}(x,\varphi)\, \sin(\varphi) \big],\\
G_{t}\,=\, -\big[ G_{1}(x,\varphi)\, \cos(3 \varphi) + G_{2}(x,\varphi)\, \sin(3 \varphi) \big], \\
G_{r}\,=\, -\big[ G_{2}(x,\varphi)\, \cos(3 \varphi) - G_{1}(x,\varphi)\, \sin(3 \varphi) \big].
\end{align}
We require that the boundary term $x\, W(x)\, \kappa(x,\varphi)$ vanishes for $x \rightarrow 0,\infty$. The derivation of $M_{\gamma}$ demands in addition that the surface terms $x W(x) \gamma_t (x, \varphi)$ vanishes for the same limits and the derivation of  $M_{\mathcal{F},\mathcal{G}}$ demands in addition to the former two that $\big[2 V(x) - x\,W(x) \big]\, \gamma_{t}(x,\varphi)$ vanishes for $x \rightarrow 0,\infty$. This can be achieved by using a suitable weighting function.\\
\\
Analogously, we derived the equivalences between the multipole moments
\begin{align}
Q^{(n)}_{\kappa}\,=\,\int_0^{\infty}\, dx\,\, x^{n+1}\,w(x)\,\int_0^{2\pi}\,d\varphi\,\, e^{i n \varphi}\, 
\kappa,\\
Q^{(n)}_{\mathcal{F}}\,=\, \int_0^{\infty}\, dx\,\, x\,W(x)\,\int_0^{2\pi}\,d\varphi\,\, e^{i n \varphi}\, \nonumber \\ (1-\kappa_{\rm rec})\, F_{t},\\
Q^{(n)}_{\gamma} \,=\, \int_0^{\infty}\, dx\,\, \big[2 W(x)\, - x^{n+1}\, w(x)\big] \int_0^{2\pi}\,d\varphi\,\, \nonumber \\
e^{i n \varphi}\, (1-\kappa_{\rm rec}) g_{t} \nonumber \\
- i\, n\, \int\limits_{0}^{\infty}\, dx\,\, W(x)\, \int\limits_{0}^{2 \pi}\, d\varphi\,\, e^{i n \varphi}\,  (1-\kappa_{\rm rec}) g_{r},\\
Q^{(n)}_{\mathcal{F},\mathcal{G}} \,=\,-\int_0^{\infty}\, dx\,\, \big[V(x)\, - \frac{1}{2} x\, W(x) \big] \int_0^{2\pi}\,d\varphi\,\, \nonumber \\
e^{i n \varphi}\,  \big[ (1-\kappa_{\rm rec}) F_{t} + (1-\kappa_{\rm rec}) G_{t} \big] \nonumber \\
+ i\, \frac{n}{2}\, \int\limits_{0}^{\infty}\, dx\,\, V(x)\, \int\limits_{0}^{2 \pi}\, d\varphi\,\, e^{i n \varphi}\, \nonumber \\
\big[ (1-\kappa_{\rm rec}) F_{r} + (1-\kappa_{\rm rec}) G_{r} \big].
\end{align}
Again we require that the term $x\, W(x)\, \kappa(x,\varphi)$ vanishes for $x \rightarrow 0,\infty$. The derivation of $Q_{\gamma}$ demands in addition that the surface term $x W(x) \gamma_t (x, \varphi)$ vanishes and the derivation of  $Q_{\mathcal{F},\mathcal{G}}$ demands in addition to the former two that $\big[2 V(x) - x\,W(x) \big]\, \gamma_{t}(x,\varphi)$ and $V(x) \gamma_r(x,\varphi)$ vanish for $x \rightarrow 0,\infty$.\\
\\
In some cases, it is useful to compute the moments only on a ring which excludes the central part of the field of view, for example if we want to exclude the strong lensing area. Therefore we extended the identities shown in this section to rings. The results are given in appendix~\ref{Appendix: Non-reduced moments on rings}.

\subsection{Mass and multipole moments of reduced shear, reduced flexion, and $K$}
The derivation of the moments and their equivalences is given in appendix~\ref{Appendix: Reduced moments on rings}. The equivalent reduced moments are:
\begin{align}
M^{(n)}_K \,=\, \int_R^{\infty}\,dx\,x^{n+1}\,w(x) \int_{0}^{2\pi}\,d\varphi\, K,\\
M^{(n)}_F\,=\,\int_R^{\infty}\,dx\,x W_R(x) \int_{0}^{2\pi}\,d\varphi\, F_t,\\
M^{(n)}_g \,=\, \int_{R}^{\infty} dx\, x W_R(x) \int_{0}^{2\pi}\,d\varphi\, (g_t F_t + g_r F_r) \nonumber\\
+ \int_R^{\infty} dx\, (2 W_R(x) - x^{n+1} w(x)) \int_{0}^{2\pi} d\varphi\, g_t,\\
M^{(n)}_{F,G} \,=\, \int_{R}^{\infty} dx\, x W_R(x) \int_{0}^{2\pi}\,d\varphi\,  (g_r F_r \nonumber \\
+ \frac{1}{2}(F_t + G_t)) \nonumber\\
- \int_R^{\infty} dx\, V_R(x) \int_{0}^{2\pi} d\varphi\, (F_t + G_t - 2 g_t F_t).
\end{align}
We defined analogous to \citet{Cain2011}
\begin{align}
K \,=\, -\ln(1-\kappa)
\end{align}
and in addition
\begin{align}
x W_R(x)\,=\, \int_{R}^{x}~dy~y^{n+1} w(y),\\
V_R(x)\,=\, \int_{R}^{x}~dy~ W_R(y).
\end{align}
We choose the lower integral limit for our ring, $R$, such that $\kappa < 1~ \forall x \geq R,\forall \varphi$. Thus $K$ is well-defined and finite. We require that $x W_R(x) K(x,\varphi)$ vanishes for $x\to R, \infty$. The $M_g$ moment requires in addition that $x W_R(x) g_t(x,\varphi)$ vanishes for $x\to R,\infty$ and the $M_{F,G}$ moment demands that additionally to the two requirements also the surface term $(2 V_R(x) - x W_R(x)) g_t(x,\varphi)$ vanishes for the same limits.\\
\\
For the multipole moments, we have the following equivalent moments:
\begin{align}
Q^{(n)}_K \,=\, \int_R^{\infty}\,dx\,x^{n+1}\,w(x) \int_{0}^{2\pi}\,d\varphi\, e^{i n \varphi}\, K,\\
Q^{(n)}_F\,=\,\int_R^{\infty}\,dx\,x W_R(x) \int_{0}^{2\pi}\,d\varphi\, e^{i n \varphi}\, F_t,\\
Q^{(n)}_g \,=\, \int_{R}^{\infty} dx\, x W_R(x) \int_{0}^{2\pi}\,d\varphi\, e^{in\varphi} (g_t F_t \nonumber \\
+ g_r F_r) \nonumber\\
+ \int_R^{\infty} dx\, (2 W_R(x) - x^{n+1} w(x)) \int_{0}^{2\pi} d\varphi\,e^{in\varphi}\, g_t \nonumber \\
-i n \int_{R}^{\infty}\,dx\, W_R(x) \int_{0}^{2\pi}\, d\varphi\, e^{in\varphi} g_r,\\
Q^{(n)}_{F,G} \,=\, \int_{R}^{\infty} dx\, x W_R(x) \int_{0}^{2\pi}\,d\varphi\,e^{in\varphi} (g_t F_t \nonumber \\
+ g_r F_r)  \nonumber \\
- \int_{R}^{\infty}\,dx\,(2V_R(x) - xW_R(x)) \int_{0}^{2\pi}\,d\varphi\, e^{in\varphi} \nonumber\\
(\frac{1}{2}(F_t + G_t) - g_t F_t) \nonumber \\
+ in \int_R^{\infty} dx\, V_R(x) \int_{0}^{2\pi} d\varphi\, e^{in\varphi} \nonumber\\
(\frac{1}{2}(F_r + G_r) - g_r F_t).
\end{align}
We have the same requirement on $R$ as for the mass moments and we demand that the resulting surface term $x W_R(x) K(x,\varphi)$ vanishes for $x\to R, \infty$. For the $Q_g$ moment we require additionally that $x W_R(x) g_t(x,\varphi)$ vanishes for the same limits. The $Q_{F,G}$ moment requires in addition to the two that the terms $(2V_R(x) - xW_R(x)) g_t(x,\varphi)$ and $V_R(x) g_r(x,\varphi)$ vanish for $x\to R$ and $x\to \infty$.\\
\\
We note that the $g$ moments require additional terms involving the flexion, which is not the case for the $\gamma$ moments. Similarly, the $F,G$ moments require shear information, which is not required for the $\mathcal{F},\mathcal{G}$ moments. 

\subsection{Mass moment transformations}

\subsubsection{Mass moments of shear, flexion, and convergence}

The mass sheet degeneracy transformations in equation~(\ref{Intro_transforms}) destroy the mass moment equivalences shown in the previous section. They hold only for $\lambda = 1$. The $n$-th order mass moment of the convergence for the transformed field $\kappa'$ is
\begin{eqnarray}
M^{(n)}_{\kappa^{'}}\,=\,\int_0^{\infty} dx\,\, x^{n+1} w(x)
\int_0^{2\pi} d\varphi\,({\lambda \kappa} + (1 - \lambda))
\end{eqnarray}
and can be written explicitly in terms of the corresponding $n$-th order
mass moments of $\kappa$,
\begin{eqnarray}
M^{(n)}_{\kappa^{'}}\,= \lambda M^{(n)}_{\kappa}\,+\,2\,\pi(1 - \lambda)\,
\int_0^{\infty}\,dx\,\, x^{n+1}\,w(x). \label{equation_kappaTransform}
\end{eqnarray}
The integral on the right-hand side of the equation above is a 
$\Gamma$ function for the choice of a Gaussian window function,  
\begin{eqnarray} 
w(x) \,=\, \frac{1}{\sqrt{2 \pi}}\, e^{-\frac{x^{2}}{2 \sigma^{2}}},
\end{eqnarray}
therefore we have
\begin{eqnarray}
M^{(n)}_{\kappa^{'}}\,= {\lambda} M^{(n)}_{\kappa} + 2^{\frac{n+1}{2}} \sqrt{\pi}\, (1-\lambda)\, \sigma^{n+2}\, \Gamma(1+\frac{n}{2}).
\end{eqnarray}
Now let us examine the transformation properties for the mass moments
computed using the shear field $\gamma$. In the limit that
$\lambda$ goes to unity, $M^{(n)}_{\kappa^{'}}$ is equivalent to 
$M^{(n)}_{\gamma^{'}}$. However, we show here that this is not the case in general, since the surface terms do not vanish identically. The moments for a transformed shear field $\gamma_t' \,=\, g_t\, (1-\kappa_{rec}') = \lambda\, g_t\, (1-\kappa_{rec}) = \lambda\, \gamma_t$ are
\begin{eqnarray}
M^{(n)}_{\gamma^{'}} &\,=\,& \int_0^{\infty}\, dx\,\, \big[2\,W(x)-x^{n+1}\,w(x)\big] \int_0^{2\pi}\, d\varphi\,\, \gamma_{t}^{'} \nonumber \\
&\,=\,& \lambda\,M^{(n)}_{\gamma}.
\end{eqnarray}
Similarly, the mass moments in terms of flexion and convergence are equivalent in the limit $\lambda \rightarrow 1$, but they differ in general. For the flexion fields $\mathcal{F}_t' \,=\, F_t\, (1-\kappa_{rec}') = \lambda\, \mathcal{F}_t$ and $\mathcal{G}_t' \,=\, G_t\, (1-\kappa_{rec}') = \lambda\, \mathcal{G}_t$ we have
\begin{eqnarray}
M^{(n)}_{\mathcal{F}^{'}} &\,=\,& \int_0^{\infty}\, dx\,\, x\,W(x)\,\int_0^{2\pi}\,d\varphi\,\, \mathcal{F}_{t}^{'} \nonumber \\
&\,=\,& \lambda\, M^{(n)}_{\mathcal{F}}, \\
M^{(n)}_{\mathcal{F}^{'},\mathcal{G}^{'}} &\,=\,& -\int_0^{\infty}\, dx\,\, \big[V(x)\, \nonumber \\
&& - \frac{1}{2} x\, W(x) \big] \int_0^{2\pi}\,d\varphi\,\, \big[ \mathcal{F}_{t}^{'} + \mathcal{G}_{t}^{'} \big] \nonumber \\
&\,=\,& \lambda\, M^{(n)}_{\mathcal{F},\mathcal{G}}.
\end{eqnarray}

Therefore, by comparing the mass moments of a given order derived from the
convergence field with that derived from the shear or flexion fields, $\lambda$
can be estimated and hence the mass degeneracy can be lifted. A comparison of the shear and flexion mass moments cannot be used to infer $\lambda$, because they are equivalent for every $\lambda$.\\
\\
As shown in appendix~\ref{Appendix:Non-reduced moments of the full field of view}, the requirements on the weighting function for the use of $M^{(n)}_{\mathcal{F},\mathcal{G}}$ are stricter than those for $M^{(n)}_{\mathcal{F}}$. The application of the former demands in the case of a simple Gaussian weighting function that the tangential shear decreases faster than $1/\log(x)$ for $x \rightarrow \infty$ to ensure that equation~(\ref{Req_mass_mom_F_G}) holds, while the use of the latter does not require such constraints. Shear typically decreases as $1/x$ and both identities are therefore usable, but it can be advantageous to design a better suited weighting function.

\subsubsection{Mass moments of reduced shear, reduced flexion, and $K$}

The mass sheet degeneracy transformations also break the equivalences of the mass moments of the reduced quantities. We have
\begin{eqnarray}
K' \,=\, -\ln(1 - (\lambda \kappa + 1 - \lambda)) \nonumber \\
\,=\, -\ln(\lambda) - \ln(1-\kappa)
\end{eqnarray}
and therefore the corresponding mass moment transforms as
\begin{eqnarray}
M^{(n)}_{K'} \,=\, \int_R^{\infty}\,dx\,x^{n+1}\,w(x) \int_{0}^{2\pi}\,d\varphi\, \nonumber\\
\big[-\ln(\lambda) - \ln(1-\kappa)\big]\\
\,=\, M^{(n)}_{K} - 2\pi \ln(\lambda) \int_R^{\infty}\,dx\,x^{n+1}\,w(x).\label{K_transform_2pi}
\end{eqnarray}
For a Gaussian weighting function, this leads to
\begin{eqnarray}
M^{(n)}_{K'}  =  M^{(n)}_{K} - 2^{\frac{n+1}{2}} \sqrt{\pi} \ln(\lambda) \sigma^{n+2} \Gamma(1 + \frac{n}{2},\frac{R^2}{2 \sigma^2}),
\end{eqnarray}
where $\Gamma(a,x)$ is the incomplete upper gamma function,
\begin{eqnarray}
\Gamma(a,x) \,=\, \int_{x}^{\infty}~dt~t^{a-1}~e^{-t}.
\end{eqnarray}
The remaining mass moments depend only on reduced quantities which do not change under this type of transformation, thus we have
\begin{eqnarray}
M^{(n)}_{g'} \,=\, M^{(n)}_g,\\
M^{(n)}_{F'} \,=\, M^{(n)}_F,\\
M^{(n)}_{F',G'} \,=\, M^{(n)}_{F,G}.
\end{eqnarray}
As we showed in the previous section, for $\lambda \to 1$ all moments of a given order are equal and we can therefore compare one of the latter three to the $K$ mass moment to determine $\lambda$.\\
\\
As seen in appendix~\ref{Appendix: Reduced moments on rings}, the requirements for the use of $M^{(n)}_{F,G}$ are again stricter than those for $M^{(n)}_F$. For a Gaussian weighting function, the reduced tangential shear has to decrease faster than $1/\log(x)$ for $x\to \infty$, which is typically the case. Designing a faster declining weight can avoid the requirement on the reduced shear.

\subsection{Multipole moment transformations}

\subsubsection{Multipole moments of shear, flexion, and convergence}

We show that the multipole moments are equivalent under the mass sheet
transformations for all $\lambda$. Therefore they cannot be used to determine its value. This is due to the
the additional exponential factor $e^{i n \varphi}$. For the following calculation, we will assume $n \geq 1$, as the multipole moments are identical to the mass moments for $n=0$. We have
\begin{eqnarray}
Q^{(n)}_{\kappa^{'}} \,=\, \nonumber \\
\int_0^{\infty}\,dx\,\, x^{n+1}\,w(x)\,\int_0^{2\pi}\,
d\varphi\,\, e^{i n \varphi}\, \big[ \lambda\,\kappa + (1 - \lambda) \big] \nonumber \\
\,=\, \lambda\, Q^{(n)}_{\kappa}.
\end{eqnarray}
The shear and flexion multipole moments transform like their mass moments,
\begin{eqnarray}
Q^{(n)}_{\gamma^{'}} &\,=\,& \lambda\, Q^{(n)}_{\gamma}, \\
Q^{(n)}_{\mathcal{F}^{'}} &\,=\,& \lambda\, Q^{(n)}_{\mathcal{F}}, \\
Q^{(n)}_{\mathcal{F}^{'},\mathcal{G}^{'}} &\,=\,& \lambda\, Q^{(n)}_{\mathcal{F},\mathcal{G}}.
\end{eqnarray}

\subsubsection{Multipole moments of reduced shear, reduced flexion, and $K$}

The multipole moments obtained from the reduced fields are also equivalent for every choice of $\lambda$ and thus they cannot be used to break the mass sheet degeneracy. As before, we restrict ourselves to the multipole moments with $n \geq 1$, because for $n=0$ they are identical to the mass moments. The $K$ multipole moment transforms as
\begin{eqnarray}
Q^{(n)}_{K'} \,=\, \int_R^{\infty}\,dx\,x^{n+1}\,w(x) \int_{0}^{2\pi}\,d\varphi\, e^{i n \varphi}\, \nonumber \\
\big[-\ln(\lambda) - \ln(1 - \kappa)\big] \nonumber\\
\,=\, Q^{(n)}_{K}.
\end{eqnarray}
Like the mass moments, the remaining multipole moments do not change under a transformation of the mass-sheet:
\begin{eqnarray}
Q^{(n)}_{g'} \,=\, Q^{(n)}_g,\\
Q^{(n)}_{F'} \,=\, Q^{(n)}_F,\\
Q^{(n)}_{F',G'} \,=\, Q^{(n)}_{F,G}.
\end{eqnarray}

\section{Application to a SIS model}

We apply our theoretical framework to a singular isothermal sphere (SIS) model to illustrate the method. The projected surface mass density is given by \citep{Bartelmann2001}:
\begin{eqnarray}
\Sigma_{\rm SIS}(\xi) \,=\, \frac{\sigma_{\nu}^{2}}{2 G \xi},
\end{eqnarray}
where $\xi$ is the separation from the lens center in the projected lens plane and $\sigma_{\nu}$ is the line-of-sight velocity dispersion of the particles. The convergence is given by $\kappa(\xi) = \Sigma(\xi)/\Sigma_{\rm crit}$ and defining the angular distance $\theta \,=\, \xi/D_{\rm ol}$ and the Einstein deflection angle
\begin{eqnarray}
\theta_{\rm E} \,=\, 4 \pi \Big(\frac{\sigma_{\nu}}{c}\Big)^{2}\, \frac{D_{\rm ls}}{D_{\rm os}}
\end{eqnarray}
leads to
\begin{eqnarray}
\kappa_{\rm SIS}(\boldsymbol{\theta}) \,=\, \frac{\theta_{\rm E}}{2 \theta}.
\end{eqnarray}
The shear for a SIS at a vectorial angular separation $\boldsymbol{\theta}$ is \citep{Bartelmann2001}
\begin{eqnarray}
\gamma_{\rm SIS}(\boldsymbol{\theta}) \,=\, - \frac{\theta_{\rm E}}{2 \theta}\, e^{2 i \varphi}
\end{eqnarray}
and for flexion, we have \citep{Bacon2006}
\begin{eqnarray}
\mathcal{F}_{\rm SIS}(\boldsymbol{\theta}) \,=\, - \frac{\theta_{\rm E}}{2 \theta^{2}} e^{i \varphi},\,\, \mathcal{G}_{\rm SIS}(\boldsymbol{\theta}) \,=\, 
\frac{3 \theta_{\rm E}}{2 \theta^{2}} e^{3 i \varphi},
\end{eqnarray}
where we used the complex shear notation $\gamma \,=\, \gamma_{1} + i\, \gamma_{2}$ and the analogous notation for flexion. Therefore the tangential and radial components are
\begin{eqnarray}
\gamma_{t, \rm SIS}(\boldsymbol{\theta}) \,=\, \frac{\theta_{\rm E}}{2 \theta},\,\, \gamma_{r, \rm SIS}(\boldsymbol{\theta}) \,=\, 0, \\
\mathcal{F}_{t, \rm SIS}(\boldsymbol{\theta}) \,=\, \frac{\theta_{\rm E}}{2 \theta^{2}},\,\, \mathcal{F}_{r, \rm SIS}(\boldsymbol{\theta}) \,=\, 0, \\
\mathcal{G}_{t, \rm SIS}(\boldsymbol{\theta}) \,=\, -\frac{3 \theta_{\rm E}}{2 \theta^{2}},\,\, \mathcal{G}_{r, \rm SIS}(\boldsymbol{\theta}) \,=\, 0.
\end{eqnarray}\\

Measuring higher order mass moments tends to be a noisy affair,
therefore we examine for simplicity the lowest order mass moment
$M^{(0)}$. Since the 0-th order mass and multipole moments are identical, we will also restrict our treatment to the mass moment. We investigate the non-reduced moment method, as these moments give short analytic expressions and the reduced moment technique is exactly analogous. Writing the tangential components in terms of $\theta$ and performing the integrals using the Gaussian weighting function, we have for the SIS
\begin{eqnarray}
M^{(0)}_{\kappa, \rm SIS} = M^{(0)}_{\gamma, \rm SIS} = M^{(0)}_{\mathcal{F}, \rm SIS} =  M^{(0)}_{\mathcal{F}, \mathcal{G}, \rm SIS} = \frac{\pi\, \sigma\, \theta_{\rm E}}{2}.\, 
\end{eqnarray} 
For the transformed moments, we have
\begin{eqnarray}
M^{(0)}_{\kappa^{'}, \rm SIS} \,=\, \lambda\, \frac{\pi\, \sigma\, \theta_{\rm E}}{2} + \sqrt{2 \pi} (1 - \lambda) \sigma^2, \\
M^{(0)}_{\gamma^{'}, \rm SIS} \,=\, M^{(0)}_{\mathcal{F}^{'}, \rm SIS} \,=\, M^{(0)}_{\mathcal{F}^{'}, \mathcal{G}^{'}, \rm SIS} \,=\, \lambda \frac{\pi\,\sigma\, \theta_{\rm E}}{2}.
\end{eqnarray}
Therefore we can estimate the value of $(1 - \lambda)$ by evaluating $\big( M^{(0)}_{\kappa^{'}, \rm SIS} - M^{(0)}_{x^{'}, \rm SIS} \big) /\sqrt{2 \pi} \sigma^2$, where $x$ represents a shear or flexion mass moment.

\section{Detectability}

In this section, we estimate the reliability of $\lambda$ detections from the mass moment measurements \textit{without assuming a halo model}. For this purpose, we compare the moments with the moment standard deviations in the absence of lensing. The latter are calculated in appendix~\ref{Appendix:Moment dispersions}. We assume that the convergence information is derived from number counts exploiting the magnification bias and that the unlensed number counts follow a power law $n_0 \propto S^{-\beta}$ with flux limit $S$ and $\beta = 0.5$ \citep[e.g., ][]{Schneider2000}. The results are
\begin{align}
\sigma_{M,\kappa} = & \sqrt{\frac{2\pi}{n_{\kappa}}} \Bigg( \int\limits_{0}^{\infty}~dx~x^{2n+1} w(x)^2 \Bigg)^{\frac{1}{2}},\\
\sigma_{M,\gamma} = & \sqrt{\frac{2\pi}{n_{\gamma}}} \sigma_{\gamma} \Bigg(\int\limits_{0}^{\infty}~dx~\frac{1}{x} [2W(x) - x^{n+1} w(x)]^2 \Bigg)^{\frac{1}{2}},\\
\sigma_{M,\mathcal{F}} = & \sqrt{\frac{2\pi}{n_{\mathcal{F}}}} \sigma_{\mathcal{F}} \Bigg(\int\limits_{0}^{\infty}~dx~xW(x)^2\Bigg)^{\frac{1}{2}},\\
\sigma_{M,\mathcal{F},\mathcal{G}} = & \sqrt{2\pi} \sqrt{\frac{\sigma_{\mathcal{F}}^2}{n_{\mathcal{F}}} + \frac{\sigma_{\mathcal{G}}^2}{n_{\mathcal{G}}}}\nonumber \\ 
& \Bigg(\int\limits_{0}^{\infty}~dx~\frac{1}{x} [V(x) - \frac{1}{2} x W(x)]^2 \Bigg)^{\frac{1}{2}}
\end{align}
for the non-reduced moments and
\begin{align} 
\sigma_{M,K} = & \Bigg( \frac{2\pi}{n_{\kappa}} \int\limits_{R}^{\infty}~dx~x^{2n+1} w(x)^2  \nonumber \\
& - \frac{\pi^2}{n_{\kappa}^2 A^2} \Bigg(\int\limits_{R}^{\infty}~dx~x^{n+1} w(x) \Bigg)^2 \Bigg)^{\frac{1}{2}}, \\
\sigma_{M,g} = & \Bigg( \frac{4\pi}{n_{\mathcal{F}}} \sigma_{\mathcal{F}}^2 \sigma_{\gamma}^2 (\int\limits_{R}^{\infty}~dx~x W_R(x)^2) \nonumber \\
& + \frac{2\pi}{n_{\gamma}} \sigma_{\gamma}^2 \int\limits_{R}^{\infty}~dx~\frac{1}{x} (2 W_R(x) - x^{n+1} w(x))^2  \Bigg)^{\frac{1}{2}}, \\
\sigma_{M,F} = & \sqrt{\frac{2\pi}{n_{\mathcal{F}}}} \sigma_{\mathcal{F}} \Bigg( \int\limits_{R}^{\infty}~dx~x W_R(x)^2 \Bigg)^{\frac{1}{2}},\\
\sigma_{M,F,G} = & \sqrt{2\pi} \Bigg( (\frac{\sigma_{\gamma}^2 \sigma_{\mathcal{F}}^2}{n_{\mathcal{F}}} + \frac{\sigma_{\mathcal{F}}^2}{4 n_{\mathcal{F}}} + \frac{\sigma_{\mathcal{G}}^2}{4 n_{\mathcal{G}}}) \int\limits_{R}^{\infty}~dx~x W_R(x)^2 \nonumber \\
& - (\frac{\sigma_{\mathcal{F}}^2}{n_{\mathcal{F}}} + \frac{\sigma_{\mathcal{G}}^2}{n_{\mathcal{G}}}) \int\limits_{R}^{\infty}~dx~W_R(x) V_R(x) \nonumber \\
& + (\frac{\sigma_{\mathcal{F}}^2}{n_{\mathcal{F}}} + \frac{\sigma_{\mathcal{G}}^2}{n_{\mathcal{G}}} + 4\frac{\sigma_{\gamma}^2 \sigma_{\mathcal{F}}^2}{n_{\mathcal{F}}}) \int\limits_{R}^{\infty}~dx~\frac{1}{x} V_R(x)^2   \Bigg)^{\frac{1}{2}}
\end{align}
for the reduced moments. Here $A$ denotes our total field of view area. Using the 2 techniques described in this paper, we can measure $1-\lambda$ or $\ln(\lambda)$ and their respective detectabilities are given by:
\begin{align}
\bigg( \frac{S}{N} \bigg)_{1 - \lambda}^{(n)} \,=\, & \Big| \frac{M^{(n)}_{\kappa^{'}} - M^{(n)}_{x^{'}}}{\big(\sigma^{2}_{M^{(n)},\kappa^{'}} + \sigma^{2}_{M^{(n)},x^{'}} \big)^{\frac{1}{2}}} \Big|,\\
\bigg( \frac{S}{N} \bigg)_{\ln(\lambda)}^{(n)} \,=\, & \Big| \frac{M^{(n)}_{K'} - M^{(n)}_{X'}}{\big(\sigma^{2}_{M^{(n)},K'} + \sigma^{2}_{M^{(n)},X'} \big)^{\frac{1}{2}}} \Big| ,
\end{align}
where $x$ and $X$ represent the non-reduced and reduced shear and flexion moments used to break the degeneracy. We calculate the detectabilities for the lowest order mass moment combinations. We assume a $\lambda$ of 1.2 and that we can measure the shear in 100 sources/$\text{arcmin}^2$, as in the \textit{Hubble Frontier Fields}. We further choose $n_{\kappa} = 300/\text{arcmin}^2$, $\sigma_{\gamma} = 0.26$ \citep{Leauthaud2007}, and $n_{\mathcal{F}} = n_{\mathcal{G}} = 25/\text{arcmin}^2$. We have $\sigma_{F a_{\rm gal}} \approx \sigma_{\mathcal{F} a_{\rm gal}}$ and $\sigma_{G a_{\rm gal}} \approx \sigma_{\mathcal{G} a_{\rm gal}}$, where $a_{gal}$ is the semi-major of the lensed source, and typical values for the intrinsic flexion of $0.03$ and $0.04$ respectively \citep{Goldberg2007,Goldberg2005}. As flexion can more reliably be measured for larger sources, we assume a lower source density and $a_{gal} = 0.2''$. This leads to $\sigma_{\mathcal{F}} = 0.15~1/\text{arcsec}$ and $\sigma_{\mathcal{G}} = 0.2~1/\text{arcsec}$. For the galaxy lens, we assume an Einstein radius of $0.2''$ and for the cluster $20''$, but note that our results do not require any assumptions on the halo shape, as the difference between the mass moments solely depends on lambda and the weighting function. We choose the lower integration boundary $R$ to be the Einstein radius and for the noise integrals involving intrinsic flexion, we choose an upper integration boundary of 3 arcmin for the galaxy lens and of 30 arcmin for the cluster instead of infinity. We can do this because flexion drops off very quickly, typically as 1/$x^2$, and therefore we can assume that the flexion and thus its noise is 0 outside of a certain, generously chosen area. E.g. for an SIS, we have that the $\mathcal{F}$ and $\mathcal{G}$ signals drop to less than $10^{-5}$ 1/arcsec at these distances. If we do not make this simplification, we would integrate the noise out to infinity, as the $W$ weighting function does not drop off quickly enough for a Gaussian $w$ function, even though there is obviously no signal in this region. In addition, we assume that we have an image area of 11 $\text{arcmin}^2$, which corresponds to the field of view of the \textit{Hubble Space Telescope Advanced Camera for Surveys}.\\
\\
For the stacking calculations, we assume that we stack 1000 galaxy-scale lenses to obtain the convergence map of an averaged galaxy lens. In this case, we can compute the $\lambda$ parameter and lift the mass sheet degeneracy for the stacked, i.e. averaged, lensing galaxy.  \\
\\
Table~\ref{table:Detactability} shows the detectability of $1 - \lambda$ and $\ln(\lambda)$ for a Gaussian weighting function $w$ and several choices of $\sigma$. A larger $\sigma$ leads to a wider weighting function that encompasses more of the lensing information and thus results in a larger detectability. If the convergence and nonreduced shear moments are combined, we clearly detect $\lambda$ and we can break the mass sheet degeneracy for stacked galaxy lenses and cluster lenses. Similarly, we can estimate $\lambda$ for stacking if we combine $K$ and reduced shear or if we use the first flexion moments, even though the latter are much noisier.\\
\\
However, in several cases the detectability is too low to reliably measure $\lambda$. The reason is that these moment combinations involve the flexion and thus the weighting function $W(x) = 1/x~\int\limits_{0}^{x}~dy~y^{n+1}~w(y)$ or  $W_R(x) = 1/x~\int\limits_{R}^{x}~dy~y^{n+1}~w(y)$, respectively. For a Gaussian weight $w(x)$ and $n=0$, the integral rapidly approaches a constant value. Thus $W(x)$ and $W_R(x)$ drop off very slowly as $1/x$, while the flexion drops off as $1/x^2$. As a result, a large area which contains no signal is highly weighted. This results in a low detectability. If a faster declining $W$ function is designed, e.g. by using a $w(x)$ which becomes negative at a certain distance from the lens center, the detectability of these moments can be substantially improved. This is in particular true for the mass moments using first and second flexion, as these require in addition the $V(x)$ and $V_R(x)$ weights, which even increase with distance. The mass moments remain finite, as the flexion drops off faster than the weight increases, but a quickly decreasing weight would increase the detectability significantly.\\
\\
A second reason for using fast declining $W$ and $V$ functions is the following: In our calculation, we could safely neglect very small flexion signals without affecting the resulting mass moment difference, because we assumed a value for $\lambda$. However, depending on $\sigma$ and the mass moment in question, these very small flexion values can be amplified by the slow drop off of the weighting function and neglecting them can lead to a potentially substantial error in the mass moment estimate. \\
\\
Several optimized aperture mass weighting functions have been proposed \citep[e.g.,][]{Leonard2009,Schneider1997}, but they typically assume that the weight $w$ is compensated, i.e. $\int_{0}^{\infty} x w(x) dx = 0$. We did not impose this condition on the weighting function in our derivations, and indeed a compensated $w$ cannot be used to break the mass sheet degeneracy using the non-reduced moments of lowest order, which are expected to have the minimal noise, as the additional term in equation (\ref{equation_kappaTransform}) would vanish.\\
\\
\citet{Maturi2005} propose a non-compensated, optimized weighting function for shear that takes into account the shape of the signal and the noise power spectrum. This approach is promising, but it is complicated by the fact that for our technique the weights $w,~W,~\text{and}~V$ are related to each other. Thus it is not possible to optimize the weighting functions independently. In fact, we must optimize at least two inter-dependent weighting functions for very different signal and noise shapes simultaneously. This optimization is outside the scope of this paper and is therefore left for a future publication. \\
\\
Lensing measurements in observational data give discrete sets rather than smooth functions of observables. An application of the aperture mass technique to observational data can therefore be achieved by either binning and approximating the integral with a Riemann sum or fitting the results to a smooth function.

\begin{table}
\centering
\caption{Detectability of $1 - \lambda$ (upper part) and $\ln(\lambda)$ (lower part) for different moment combinations. The results are shown for galaxy lenses, 1000 stacked galaxy lenses, and cluster lenses for different $\sigma$ values of the Gaussian $w(x)$ weighting function. The moment combinations listed after the first one involve the $W$ or $V$ weighting function types, which drop off very slowly. Thus their detectability can be substantially improved by designing a faster declining weighting function (see text).   }\label{table:Detactability}

\resizebox{\columnwidth}{!}{
\begin{tabular}{|l|l|l|l|l|l|l|l|l|l|}
\hline
& \multicolumn{3}{c|}{Galaxy lensing} & \multicolumn{3}{c|}{Stacking} & \multicolumn{3}{c|}{Cluster lensing} \\
\hline
$\sigma$ & $2''$ & $5''$ & $10''$ & $2''$ & $5''$ & $10''$ & $20''$ & $50''$ & $100''$\\
\hline
$\kappa + \gamma$ & 0.4 & 0.9 & 1.9 & 11.8 & 29.5 & 59.0 & 3.7 & 9.3 & 18.7 \\
$\kappa + \mathcal{F}$ & 0.1 & 0.2 & 0.2 & 4.1 & 4.9 & 5.6 & 0.1 & 0.2 & 0.2\\
$\kappa + \mathcal{F},\mathcal{G}$  & 0.04 & 0.05 & 0.08 & 1.2 & 1.7 & 2.6 & 0.04 & 0.05 & 0.08 \\
\hline
$K + g$  & 0.2 & 0.4 & 0.4 & 7.6 & 11.1 & 13.4 & 0.3 & 0.4 & 0.4 \\
$K + F$  & 0.1 & 0.1 & 0.2 & 3.8 & 4.5 & 5.1 & 0.1 & 0.1 & 0.2 \\
$K + F,G$  & 0.03 & 0.05 & 0.07 & 1.0 & 1.5 & 2.1 & 0.04 & 0.05 & 0.07\\
\hline
\end{tabular}
}
\end{table}

\section{Convergence, shear, and flexion signal-to-noise}
 
\begin{figure}
\capstart
\includegraphics[width = \columnwidth]{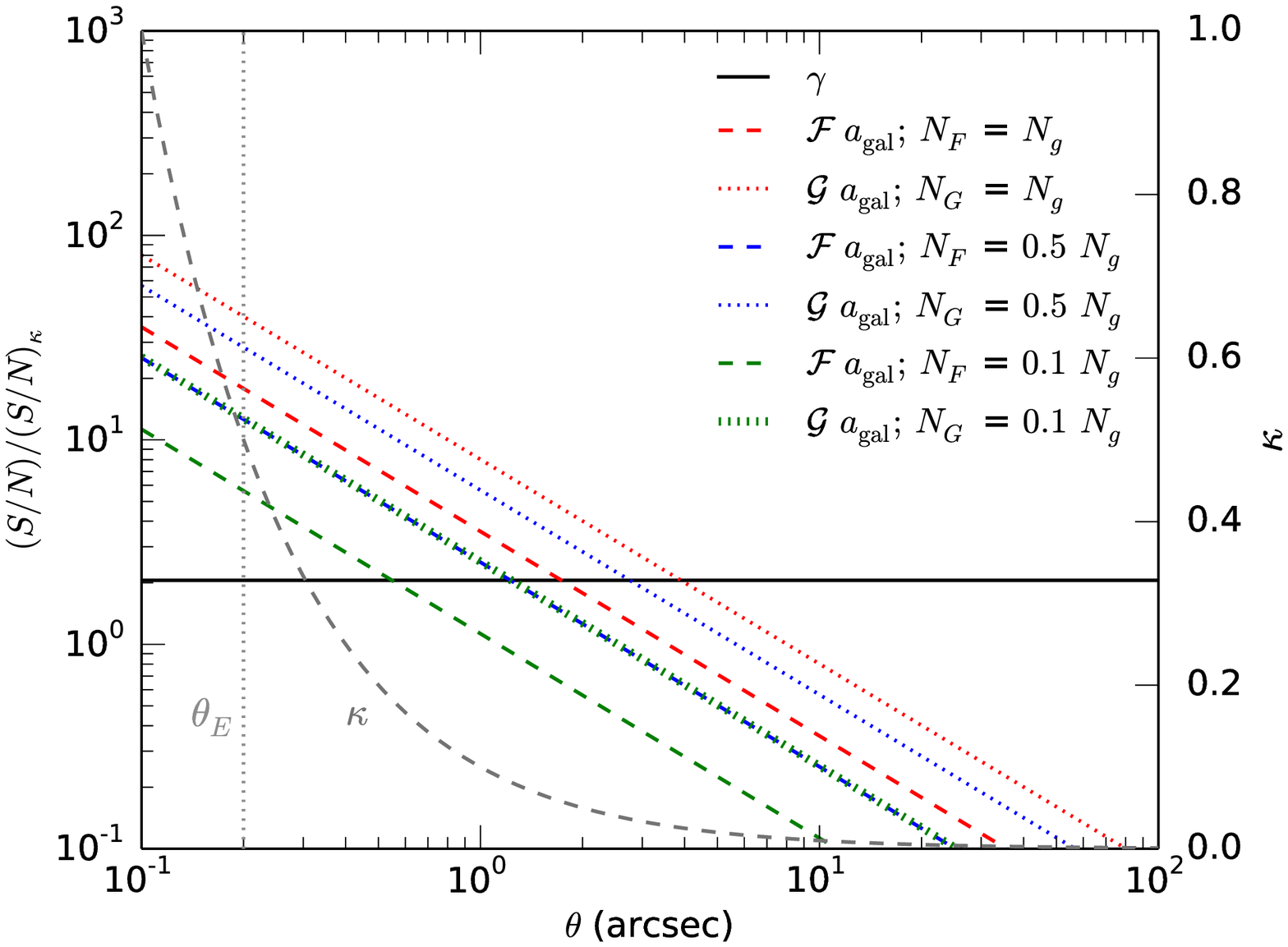}
\includegraphics[width = \columnwidth]{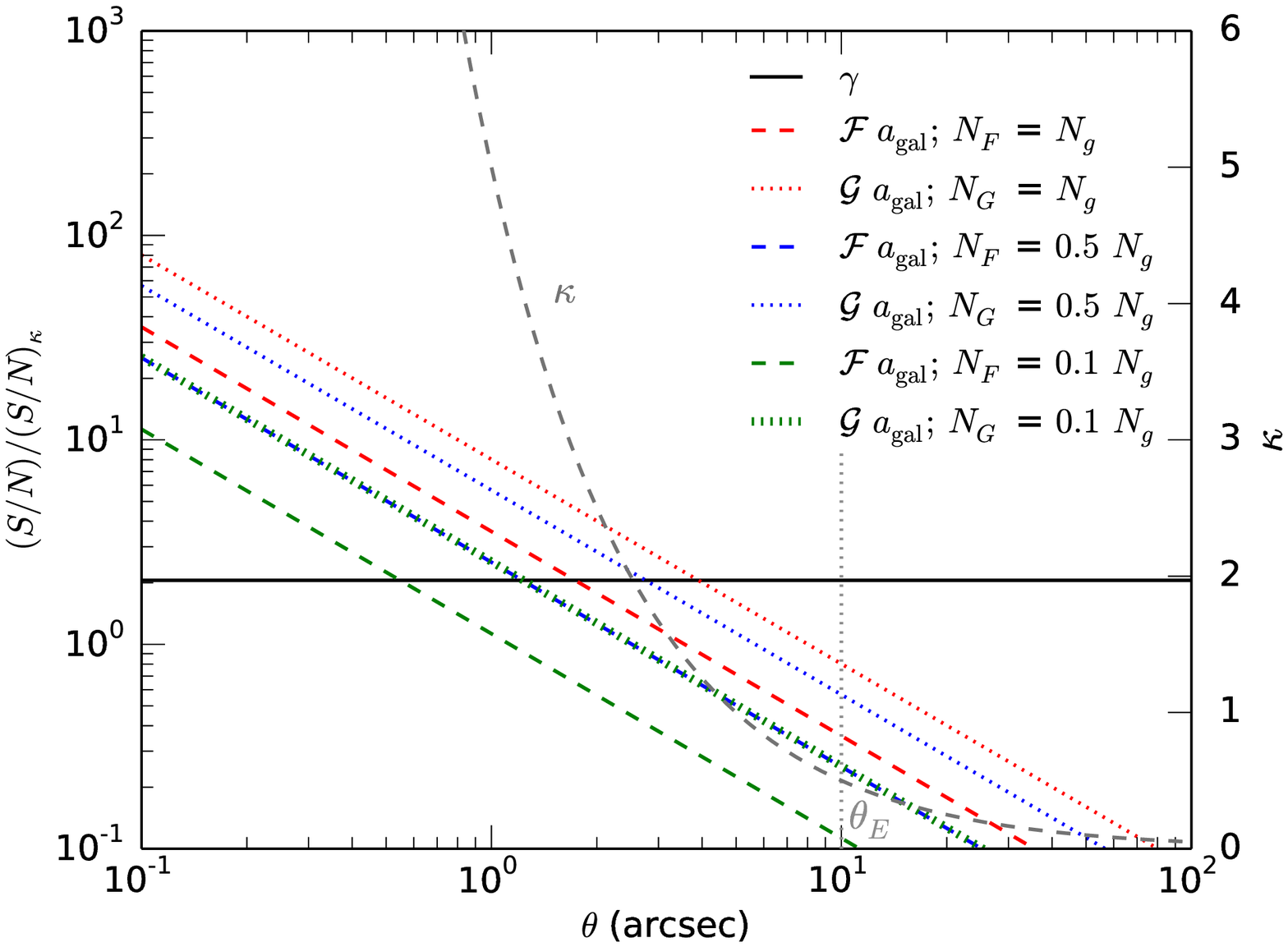}
\includegraphics[width = \columnwidth]{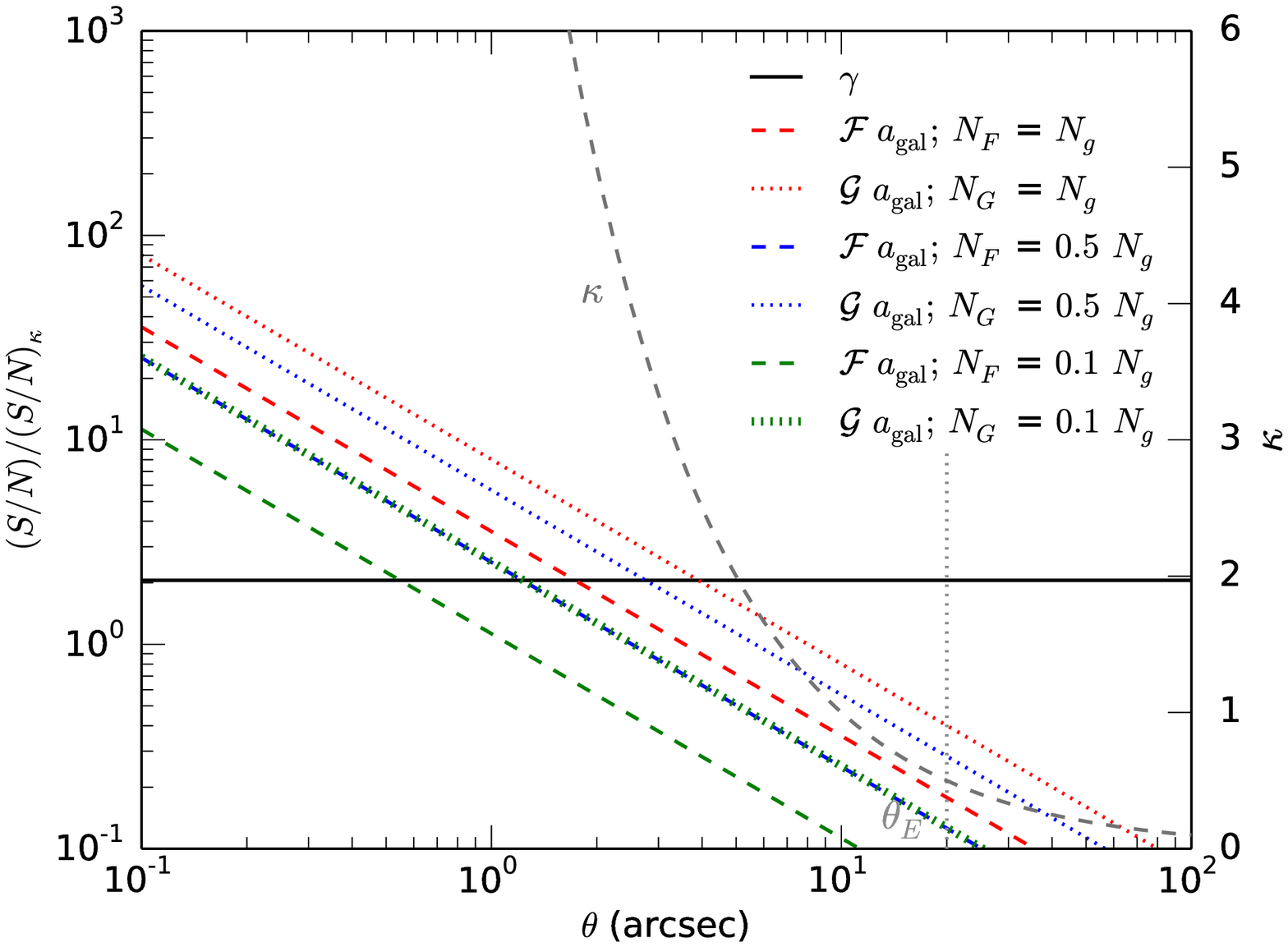}
\caption{Comparison of the signal-to-noise ratios of $\gamma$, $\mathcal{F}\, a_{\rm gal}$, and $\mathcal{G}\, a_{\rm gal}$ normalized to $(S/N)_\kappa$, the signal-to-noise of number counts, for the singular isothermal sphere. The Einstein radii are $0.2''$ (top), $10''$ (middle), and $20''$ (bottom). We show the flexion results for source densities of 100\%, 50\%, and 10\% of the shear source density and plot the convergence value as a reference.} \label{Fig_1}
\end{figure}

\begin{figure}
\capstart
\includegraphics[width = \columnwidth]{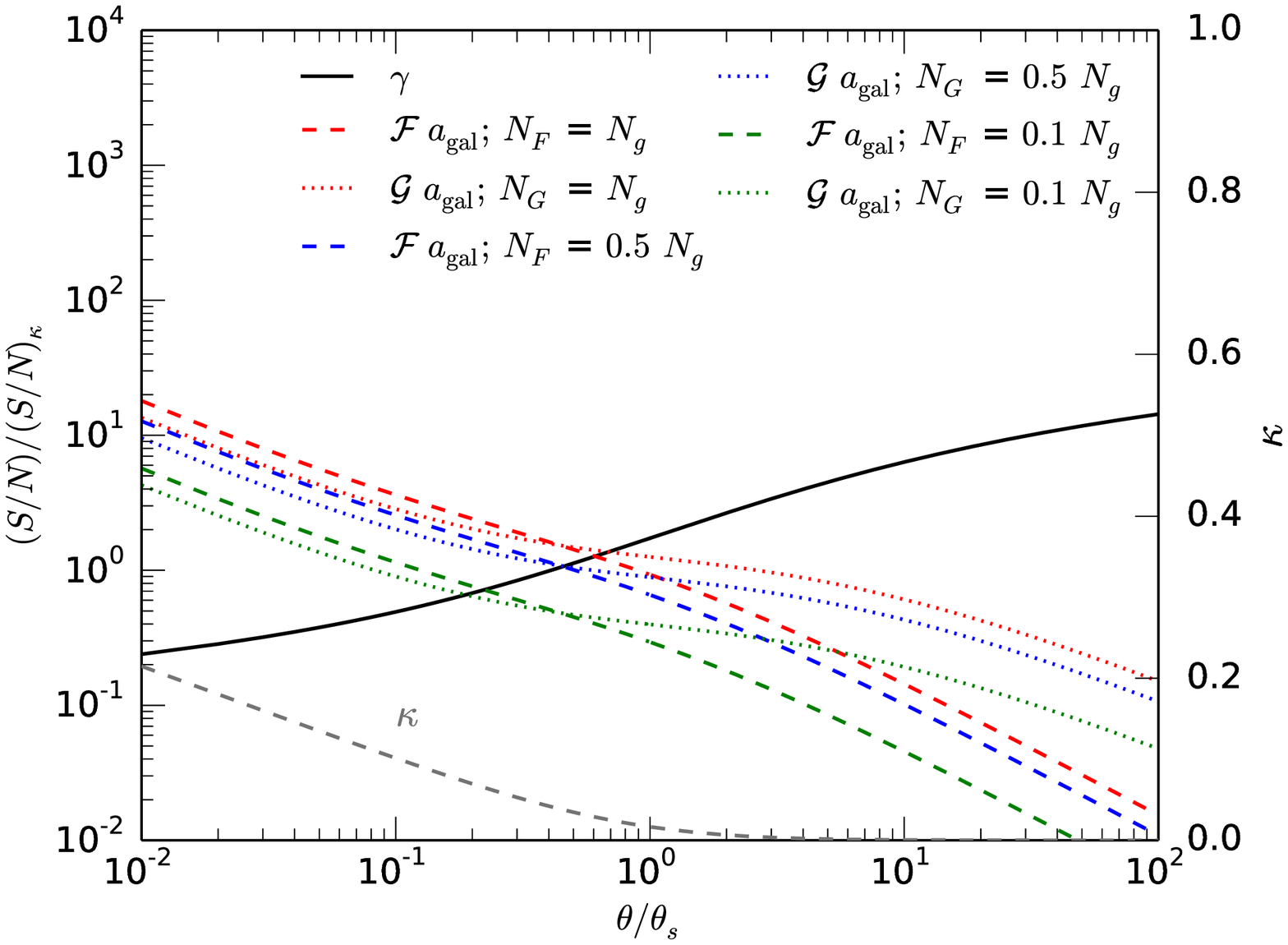}
\includegraphics[width = \columnwidth]{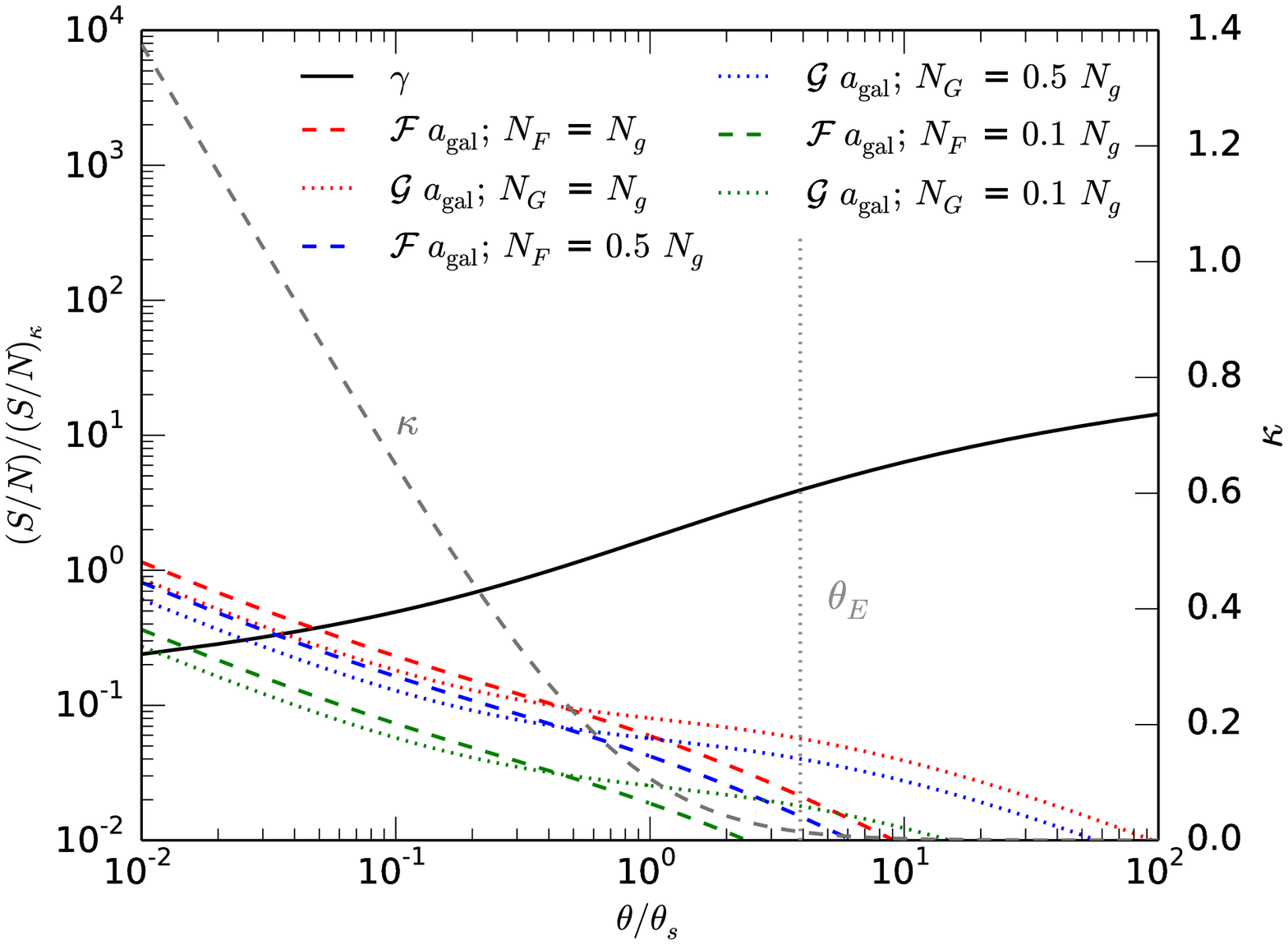}
\includegraphics[width = \columnwidth]{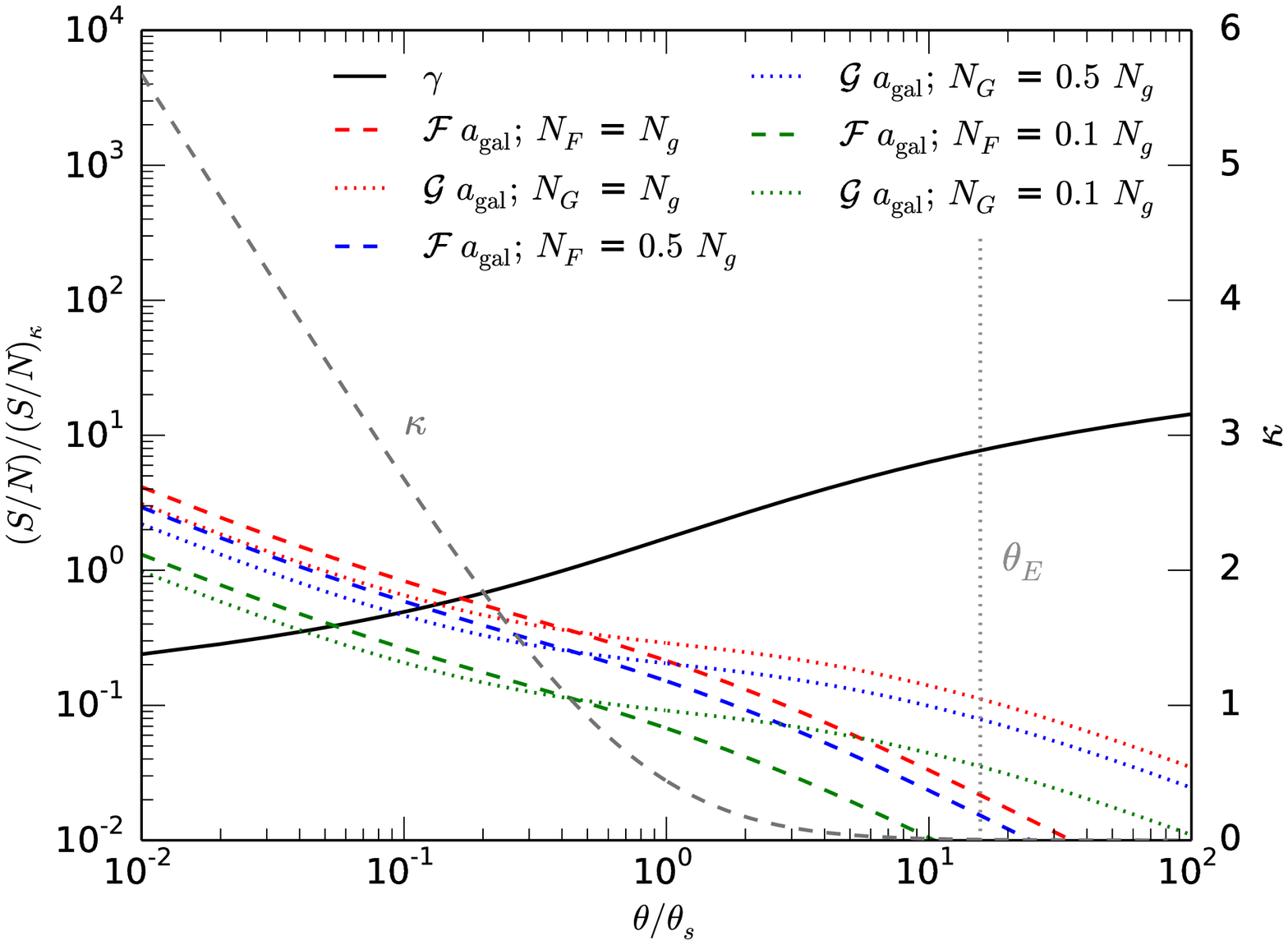}
\caption{Comparison of $(S/N)$ ratios of $\gamma$, $\mathcal{F}\, a_{\rm gal}$, and $\mathcal{G}\, a_{\rm gal}$ normalized to $(S/N)_{\kappa}$, the signal-to-noise of number counts, for NFW profiles with $\theta_{s} \,=\, 4.6 \arcsec$ (top), $72 \arcsec$ (middle), and $20 \arcsec$ (bottom). The flexion is shown for source densities of 100\%, 50\%, and 10\% of the shear source density. The convergence and the Einstein radius are plotted as a reference. The Einstein radius of the galaxy is too small to be shown.} \label{Fig_2}
\end{figure}

We compare the signal-to-noise ratios for SIS and Navarro-Frenk-White (NFW) profiles \citep{Navarro1996,Navarro1997}. We show that the combination of shear and flexion has a higher signal-to-noise ratio than the number counts method for both galaxy and cluster lensing. Therefore including number counts in the mass sheet reconstruction does not improve the lens model. Thus we can use this information to break the mass sheet degeneracy without sacrificing the accuracy of the model.\\
\\
We can restrict ourselves to the non-reduced shear and flexion, because the extra $1/(1-\kappa)$ term in the reduced quantities would only boost their signal-to-noise. The convergence, shear, and flexion formulas for a NFW profile are given in appendix~\ref{Appendix: NFW formulas}. In addition, we only investigate the intrinsic noise of each method and neglect possible additional measurement errors.\\
\\
We can make flexion dimensionless by multiplying it with the semi-major of the observed galaxy $a_{\rm gal}$ \citep{Goldberg2005}. The signal-to-noise ratios are
\begin{eqnarray}
\bigg( \frac{S}{N} \bigg)_{F a_{\rm gal}} \,=\, \frac{|F|\, a_{\rm gal}\, \sqrt{N_{F}}}{\sigma_{F a_{\rm gal}}}, \\
\bigg( \frac{S}{N} \bigg)_{G a_{\rm gal}} \,=\, \frac{|G|\, a_{\rm gal}\, \sqrt{N_{G}}}{\sigma_{G a_{\rm gal}}},
\end{eqnarray}
where $\sigma_{F a_{\rm gal}}$ and $\sigma_{G a_{\rm gal}}$ are the dispersions for dimensionless reduced flexion and we average over $N_F$ and $N_G$ sources. We can infer the signal-to-noise of $\mathcal{F}\, a_{\rm gal}$ and $\mathcal{G}\, a_{\rm gal}$ by replacing the reduced quantities with these expressions. We have again $\sigma_{F a_{\rm gal}} \approx \sigma_{\mathcal{F} a_{\rm gal}}$ and $\sigma_{G a_{\rm gal}} \approx \sigma_{\mathcal{G} a_{\rm gal}}$ with typical values of $0.03$ and $0.04$ respectively \citep{Goldberg2007,Goldberg2005}. The signal-to-noise of the shear and number counts is \citep{Schneider2000}
\begin{align}
\bigg( \frac{S}{N} \bigg)_{g} \,=\, \frac{|g| \sqrt{N_g}}{\sigma_{\epsilon}},\\
\bigg( \frac{S}{N} \bigg)_{\kappa} \,=\, |\mu^{\beta - 1} - 1| \sqrt{N_{\kappa}} \approx 2\kappa |1-\beta| \sqrt{N_{\kappa}},
\end{align}
where we have used the first order weak lensing expansion. We average our number counts over $N_{\kappa}$ sources and $\beta$ is the number counts exponent, see appendix~\ref{Appendix:Moment dispersions}.
As we investigate the non-reduced quantities, we replace $g$ with $\gamma$. We choose  $\sigma_{\epsilon} \,=\, 0.26$ \citep{Leauthaud2007} and the remaining parameters analogously to \citet{Schneider2000}, $\beta \,=\, 0.5$ and $N_{\kappa} \,=\, 3.5\, N_{g}$. The latter is a slightly optimistic estimate for $N_{\kappa}$, since for real observations, we will loose several sources due to color cuts. A lower estimate would decrease the signal-to-noise of the number counts method, so we will keep this value. Using the expressions for the SIS, we have
\begin{eqnarray}
\frac{(S/N)_{\gamma}}{(S/N)_{\kappa}} &\,=\,& 2.1, \\
\frac{(S/N)_{\mathcal{F} a_{\rm gal}}}{(S/N)_{\kappa}} &\,=\,& \frac{a_{\rm gal}}{0.03\,\, \theta}\, \sqrt{\frac{N_{F}}{N_{\kappa}}}, \\
\frac{(S/N)_{\mathcal{G} a_{\rm gal}}}{(S/N)_{\kappa}} &\,=\,& \frac{3\, a_{\rm gal}}{0.04\,\, \theta}\, \sqrt{\frac{N_{G}}{N_{\kappa}}}.
\end{eqnarray}
The radial dependence of the ratios is shown in figure~\ref{Fig_1} for $a_{\rm gal} \,=\, 0.2 \arcsec$ and $N_F = N_G$ values of $N_g$, $0.5\, N_g$, and $0.1\, N_g$. Note that the signal-to-noise ratios do not depend on the Einstein radius $\theta_{\rm E}$, which is canceled in the calculation. However, the distance of the weak lensing regime from the halo center does depend on the Einstein radius. Therefore we show $\kappa$ for Einstein radii of $0.2''$, $10''$, and $20''$. For larger Einstein radii, shear by itself has always the best signal-to-noise in the weak lensing regime. For smaller $\theta_E$, shear and flexion together give the best results. Note that even in the case of large Einstein radii, the flexion information is not redundant. It is much more sensitive to small scale structures than shear and can therefore be used to reconstruct them \citep[][]{Leonard2009,Bacon2010}. \\
\\
The analytic expressions for the NFW profile are complex and therefore it is more instructing to plot the signal-to-noise ratios. These are independent of $\rho_{\rm crit}$, $\delta_{c}$, and $\Sigma_{\rm crit}$. We show the radial dependence for three angular scale radii $\theta_{s} \,=\, r_{s}/D_{\rm ol}$ in figure~\ref{Fig_2}. The first plot corresponds to a galaxy at redshift $z \,=\, 0.35$ with $r_s \,=\, 16\, h^{-1}~\kpc $ and sources at redshift $z = 0.6$, which agrees e.g. with the observations in \citet{Hoekstra2004}. The second and third correspond to clusters with $r_{s} \,=\, 250\, h^{-1}~\kpc$ and $r_{s} \,=\, 70\, h^{-1}~\kpc$  at the same redshift, comparable to the NFW models for one component of Abell 370 and for the cluster MS 2137 in \citet{Shu2008}. For the first cluster, we only model one component to obtain a NFW profile with intermediate Einstein radius for our comparison. Assuming a standard, flat $\rm \Lambda CDM$ cosmology with $\Omega_{m} \,=\, 0.3$, $\Omega_{\Lambda} \,=\, 0.7$ and $h \,=\, 0.7$, we calculate the angular diameter distances using the \citet{Wright2006} cosmology calculator and astropy \citep{AstropyCollaboration2013} and obtain the corresponding angular scale radii $ 4.6 \arcsec$, $72 \arcsec$, and $ 20 \arcsec$. For the galaxy, we further set $\delta_{c} = 2.4 \cdot 10^4$ and for the two clusters $\kappa_s = \rho_{\rm crit} \delta_c r_s / \Sigma_{\rm crit} = 0.16$ and $0.66$, in agreement with the values cited in the respective publications. The size of the semi-major and the number of measured sources are the same as for the SIS. The effect of a different choice of these parameters can be visualized by a shift of the corresponding graphs, because the scale in the figures is logarithmic. \\
\\ 
Figure~\ref{Fig_2} shows that shear by itself has always the best signal-to-noise ratio in the weak lensing regime of the 2 clusters. For the galaxy, a combination of flexion and shear provides the best results, as the flexion dominates at small separations from the center. This illustrates also the value of flexion for the reconstruction of substructures.

\section{Discussion and Conclusions}

We have derived mass and multipole moment formulas in terms of flexion and showed that they are equivalent to the mass and multipole moments of shear and convergence. Furthermore, we have showed the equivalence of mass and multipole moments in terms of reduced shear, reduced flexion, and $K$, which is a quantity derived from the convergence. We investigated the moment behavior with respect to the mass sheet degeneracy transformation and found that the equivalences are broken for the mass moments, but preserved for the multipole moments. The resulting surface terms can be used to break the mass sheet degeneracy.\\
\\
We demonstrated the new theoretical framework by applying it to a SIS mass model. In addition, we investigated the detectability of the mass sheet parameter $\lambda$ without assuming a halo model. We found that we can break the mass sheet degeneracy for stacked galaxy-galaxy and cluster lensing. Combinations of the shear and convergence moments have a much higher detectability than combinations using the flexions. This is due to their weighting function, which drops off too slowly. A fine-tuned weight will significantly improve their performance.\\
\\
Finally, we investigated the signal-to-noise ratios of shear, flexion, and convergence information from number counts for SIS and NFW halos in the weak lensing regime. It was assumed that the noise in $\gamma$, $\mathcal{F}$, and $\mathcal{G}$ is dominated by the intrinsic signal of the lensed sources and that the number counts are dominated by Poisson noise. The estimates do not include measurement errors e.g. due to pixel noise. This will have an additional effect, in particular on flexion and if sources with insufficient signal-to-noise are used \citep{Rowe2013}. With high quality data, e.g. from the \textit{Hubble Frontier Fields}, the reliability of the flexion measurements should improve. Under these assumptions, we demonstrated that the combination of shear and flexion always gives better results than other combinations using the number counts. Using an estimate of shear and number count signal-to-noise ratios, e.g. \citet{Schneider2000} also found that shear is the superior method. However, \citet{VanWaerbeke2010} and \citet{Hildebrandt2013} caution that we can still count sources at redshifts for which shear measurements are no longer feasible, thus reducing the advantage of shear. We find that the convergence information can thus be used to break the mass sheet degeneracy without a loss in the quality of the lens model.

\section*{Acknowledgements}
MR thanks the Department of Astronomy at Yale University for hosting him and acknowledges support by the German National Academic Foundation. He thanks Thibault Kuntzer, Austen Groener, and Justin Bird for fruitful discussions and Alexandre Refregier for his support. MR and PN acknowledge useful discussions with David Goldberg and Richard Massey on flexions. MR and JPK acknowledge support from the ERC advanced grant LIDA. PN gratefully acknowledges support from an NSF theory program via the grant AST-1044455 and a theory grant from the Space Telescope Science Institute HST-AR-12144.01-A.

\bibliography{Astronomy_papers.bib}

\begin{thebibliography}{}
\makeatletter
\relax
\def\mn@urlcharsother{\let\do\@makeother \do\$\do\&\do\#\do\^\do\_\do\%\do\~}
\def\mn@doi{\begingroup\mn@urlcharsother \@ifnextchar [ {\mn@doi@}
  {\mn@doi@[]}}
\def\mn@doi@[#1]#2{\def\@tempa{#1}\ifx\@tempa\@empty \href
  {http://dx.doi.org/#2} {doi:#2}\else \href {http://dx.doi.org/#2} {#1}\fi
  \endgroup}
\def\mn@eprint#1#2{\mn@eprint@#1:#2::\@nil}
\def\mn@eprint@arXiv#1{\href {http://arxiv.org/abs/#1} {{\tt arXiv:#1}}}
\def\mn@eprint@dblp#1{\href {http://dblp.uni-trier.de/rec/bibtex/#1.xml}
  {dblp:#1}}
\def\mn@eprint@#1:#2:#3:#4\@nil{\def\@tempa {#1}\def\@tempb {#2}\def\@tempc
  {#3}\ifx \@tempc \@empty \let \@tempc \@tempb \let \@tempb \@tempa \fi \ifx
  \@tempb \@empty \def\@tempb {arXiv}\fi \@ifundefined
  {mn@eprint@\@tempb}{\@tempb:\@tempc}{\expandafter \expandafter \csname
  mn@eprint@\@tempb\endcsname \expandafter{\@tempc}}}

\bibitem[\protect\citeauthoryear{{Astropy Collaboration} et~al.,}{{Astropy
  Collaboration} et~al.}{2013}]{AstropyCollaboration2013}
{Astropy Collaboration} et~al., 2013, \mn@doi [\aap]
  {10.1051/0004-6361/201322068}, \href
  {http://adsabs.harvard.edu/abs/2013A%26A...558A..33A} {558, A33}

\bibitem[\protect\citeauthoryear{{Bacon}, {Goldberg}, {Rowe}  \&
  {Taylor}}{{Bacon} et~al.}{2006}]{Bacon2006}
{Bacon} D.~J.,  {Goldberg} D.~M.,  {Rowe} B.~T.~P.,   {Taylor} A.~N.,  2006,
  \mn@doi [\mnras] {10.1111/j.1365-2966.2005.09624.x}, 365, 414

\bibitem[\protect\citeauthoryear{{Bacon}, {Amara}  \& {Read}}{{Bacon}
  et~al.}{2010}]{Bacon2010}
{Bacon} D.~J.,  {Amara} A.,   {Read} J.~I.,  2010, \mn@doi [\mnras]
  {10.1111/j.1365-2966.2010.17316.x}, 409, 389

\bibitem[\protect\citeauthoryear{{Bartelmann}}{{Bartelmann}}{1996}]{Bartelmann1996}
{Bartelmann} M.,  1996, \aap, 313, 697

\bibitem[\protect\citeauthoryear{{Bartelmann} \& {Narayan}}{{Bartelmann} \&
  {Narayan}}{1995}]{Bartelmann1995}
{Bartelmann} M.,  {Narayan} R.,  1995, \mn@doi [\apj] {10.1086/176200}, 451, 60

\bibitem[\protect\citeauthoryear{{Bartelmann} \& {Schneider}}{{Bartelmann} \&
  {Schneider}}{2001}]{Bartelmann2001}
{Bartelmann} M.,  {Schneider} P.,  2001, \mn@doi [\physrep]
  {10.1016/S0370-1573(00)00082-X}, 340, 291

\bibitem[\protect\citeauthoryear{{Bartelmann}, {Narayan}, {Seitz}  \&
  {Schneider}}{{Bartelmann} et~al.}{1996}]{Bartelmann1996a}
{Bartelmann} M.,  {Narayan} R.,  {Seitz} S.,   {Schneider} P.,  1996, \mn@doi
  [\apjl] {10.1086/310114}, 464, L115

\bibitem[\protect\citeauthoryear{{Brada{\v c}}, {Lombardi}  \&
  {Schneider}}{{Brada{\v c}} et~al.}{2004}]{Bradavc2004}
{Brada{\v c}} M.,  {Lombardi} M.,   {Schneider} P.,  2004, \mn@doi [\aap]
  {10.1051/0004-6361:20035744}, 424, 13

\bibitem[\protect\citeauthoryear{{Broadhurst}, {Taylor}  \&
  {Peacock}}{{Broadhurst} et~al.}{1995}]{Broadhurst1995}
{Broadhurst} T.~J.,  {Taylor} A.~N.,   {Peacock} J.~A.,  1995, \mn@doi [\apj]
  {10.1086/175053}, 438, 49

\bibitem[\protect\citeauthoryear{{Cain}, {Schechter}  \& {Bautz}}{{Cain}
  et~al.}{2011}]{Cain2011}
{Cain} B.,  {Schechter} P.~L.,   {Bautz} M.~W.,  2011, \mn@doi [\apj]
  {10.1088/0004-637X/736/1/43}, \href
  {http://adsabs.harvard.edu/abs/2011ApJ...736...43C} {736, 43}

\bibitem[\protect\citeauthoryear{{Dye} et~al.,}{{Dye} et~al.}{2002}]{Dye2002}
{Dye} S.,  et~al., 2002, \mn@doi [\aap] {10.1051/0004-6361:20020226}, \href
  {http://adsabs.harvard.edu/abs/2002A%26A...386...12D} {386, 12}

\bibitem[\protect\citeauthoryear{{Er}, {Li}  \& {Schneider}}{{Er}
  et~al.}{2010}]{Er2010}
{Er} X.,  {Li} G.,   {Schneider} P.,  2010, preprint (\mn@eprint {arXiv}
  {1008.3088})

\bibitem[\protect\citeauthoryear{{Fort}, {Mellier}  \& {Dantel-Fort}}{{Fort}
  et~al.}{1997}]{Fort1997}
{Fort} B.,  {Mellier} Y.,   {Dantel-Fort} M.,  1997, \aap, \href
  {http://adsabs.harvard.edu/abs/1997A%26A...321..353F} {321, 353}

\bibitem[\protect\citeauthoryear{{Goldberg} \& {Bacon}}{{Goldberg} \&
  {Bacon}}{2005}]{Goldberg2005}
{Goldberg} D.~M.,  {Bacon} D.~J.,  2005, \mn@doi [\apj] {10.1086/426782}, 619,
  741

\bibitem[\protect\citeauthoryear{{Goldberg} \& {Leonard}}{{Goldberg} \&
  {Leonard}}{2007}]{Goldberg2007}
{Goldberg} D.~M.,  {Leonard} A.,  2007, \mn@doi [\apj] {10.1086/513137}, 660,
  1003

\bibitem[\protect\citeauthoryear{{Goldberg} \& {Natarajan}}{{Goldberg} \&
  {Natarajan}}{2002}]{Goldberg2002}
{Goldberg} D.~M.,  {Natarajan} P.,  2002, \mn@doi [\apj] {10.1086/324202}, 564,
  65

\bibitem[\protect\citeauthoryear{Goodman}{Goodman}{1960}]{Goodman1960}
Goodman L.~A.,  1960, Journal of the American Statistical Association, 55, 708

\bibitem[\protect\citeauthoryear{{Hildebrandt} et~al.,}{{Hildebrandt}
  et~al.}{2013}]{Hildebrandt2013}
{Hildebrandt} H.,  et~al., 2013, \mn@doi [\mnras] {10.1093/mnras/sts585}, \href
  {http://adsabs.harvard.edu/abs/2013MNRAS.429.3230H} {429, 3230}

\bibitem[\protect\citeauthoryear{{Hoekstra}, {Yee}  \& {Gladders}}{{Hoekstra}
  et~al.}{2004}]{Hoekstra2004}
{Hoekstra} H.,  {Yee} H.~K.~C.,   {Gladders} M.~D.,  2004, \mn@doi [\apj]
  {10.1086/382726}, 606, 67

\bibitem[\protect\citeauthoryear{{Irwin} \& {Shmakova}}{{Irwin} \&
  {Shmakova}}{2006}]{Irwin2006}
{Irwin} J.,  {Shmakova} M.,  2006, \mn@doi [\apj] {10.1086/504100}, 645, 17

\bibitem[\protect\citeauthoryear{{Kaiser}}{{Kaiser}}{1995}]{Kaiser1995a}
{Kaiser} N.,  1995, \mn@doi [\apjl] {10.1086/187730}, 439, L1

\bibitem[\protect\citeauthoryear{{Kaiser} \& {Squires}}{{Kaiser} \&
  {Squires}}{1993}]{Kaiser1993}
{Kaiser} N.,  {Squires} G.,  1993, \mn@doi [\apj] {10.1086/172297}, 404, 441

\bibitem[\protect\citeauthoryear{{Kneib} \& {Natarajan}}{{Kneib} \&
  {Natarajan}}{2011}]{Kneib2011}
{Kneib} J.-P.,  {Natarajan} P.,  2011, \mn@doi [\aapr]
  {10.1007/s00159-011-0047-3}, 19, 47

\bibitem[\protect\citeauthoryear{{Leauthaud} et~al.,}{{Leauthaud}
  et~al.}{2007}]{Leauthaud2007}
{Leauthaud} A.,  et~al., 2007, \mn@doi [\apjs] {10.1086/516598}, 172, 219

\bibitem[\protect\citeauthoryear{{Leonard}, {Goldberg}, {Haaga}  \&
  {Massey}}{{Leonard} et~al.}{2007}]{Leonard2007}
{Leonard} A.,  {Goldberg} D.~M.,  {Haaga} J.~L.,   {Massey} R.,  2007, \mn@doi
  [\apj] {10.1086/520109}, 666, 51

\bibitem[\protect\citeauthoryear{{Leonard}, {King}  \& {Wilkins}}{{Leonard}
  et~al.}{2009}]{Leonard2009}
{Leonard} A.,  {King} L.~J.,   {Wilkins} S.~M.,  2009, \mn@doi [\mnras]
  {10.1111/j.1365-2966.2009.14546.x}, 395, 1438

\bibitem[\protect\citeauthoryear{{Maturi}, {Meneghetti}, {Bartelmann}, {Dolag}
  \& {Moscardini}}{{Maturi} et~al.}{2005}]{Maturi2005}
{Maturi} M.,  {Meneghetti} M.,  {Bartelmann} M.,  {Dolag} K.,   {Moscardini}
  L.,  2005, \mn@doi [\aap] {10.1051/0004-6361:20042600}, \href
  {http://adsabs.harvard.edu/abs/2005A%26A...442..851M} {442, 851}

\bibitem[\protect\citeauthoryear{{Meylan}, {Jetzer}  \& {North}}{{Meylan}
  et~al.}{2006}]{Schneider2006}
{Meylan} G.,  {Jetzer} P.,   {North} P.,  eds, 2006, {Gravitational Lensing:
  Strong, Weak and Micro}  (\mn@eprint {} {astro-ph/0407232}), \url
  {http://adsabs.harvard.edu/abs/2006glsw.conf.....M}

\bibitem[\protect\citeauthoryear{{Natarajan}, {Kneib}, {Smail}  \&
  {Ellis}}{{Natarajan} et~al.}{1998}]{Natarajan1998}
{Natarajan} P.,  {Kneib} J.-P.,  {Smail} I.,   {Ellis} R.~S.,  1998, \apj,
  \href {http://adsabs.harvard.edu/abs/1998ApJ...499..600N} {499, 600}

\bibitem[\protect\citeauthoryear{{Navarro}, {Frenk}  \& {White}}{{Navarro}
  et~al.}{1996}]{Navarro1996}
{Navarro} J.~F.,  {Frenk} C.~S.,   {White} S.~D.~M.,  1996, \mn@doi [\apj]
  {10.1086/177173}, 462, 563

\bibitem[\protect\citeauthoryear{{Navarro}, {Frenk}  \& {White}}{{Navarro}
  et~al.}{1997}]{Navarro1997}
{Navarro} J.~F.,  {Frenk} C.~S.,   {White} S.~D.~M.,  1997, \mn@doi [\apj]
  {10.1086/304888}, 490, 493

\bibitem[\protect\citeauthoryear{{Okura}, {Umetsu}  \& {Futamase}}{{Okura}
  et~al.}{2007}]{Okura2007}
{Okura} Y.,  {Umetsu} K.,   {Futamase} T.,  2007, \mn@doi [\apj]
  {10.1086/513135}, 660, 995

\bibitem[\protect\citeauthoryear{{Okura}, {Umetsu}  \& {Futamase}}{{Okura}
  et~al.}{2008}]{Okura2008}
{Okura} Y.,  {Umetsu} K.,   {Futamase} T.,  2008, \mn@doi [\apj]
  {10.1086/587676}, 680, 1

\bibitem[\protect\citeauthoryear{{Rexroth}}{{Rexroth}}{2015}]{Rexroth2015}
{Rexroth} M.,  2015, submitted to IAU Proceedings Astronomy in Focus

\bibitem[\protect\citeauthoryear{{Rowe}, {Bacon}, {Massey}, {Heymans},
  {H{\"a}u{\ss}ler}, {Taylor}, {Rhodes}  \& {Mellier}}{{Rowe}
  et~al.}{2013}]{Rowe2013}
{Rowe} B.,  {Bacon} D.,  {Massey} R.,  {Heymans} C.,  {H{\"a}u{\ss}ler} B.,
  {Taylor} A.,  {Rhodes} J.,   {Mellier} Y.,  2013, \mn@doi [\mnras]
  {10.1093/mnras/stt1353}, 435, 822

\bibitem[\protect\citeauthoryear{{Schneider} \& {Bartelmann}}{{Schneider} \&
  {Bartelmann}}{1997}]{Schneider1997}
{Schneider} P.,  {Bartelmann} M.,  1997, \mnras, 286, 696

\bibitem[\protect\citeauthoryear{{Schneider} \& {Er}}{{Schneider} \&
  {Er}}{2008}]{Schneider2008}
{Schneider} P.,  {Er} X.,  2008, \mn@doi [\aap] {10.1051/0004-6361:20078631},
  485, 363

\bibitem[\protect\citeauthoryear{{Schneider} \& {Seitz}}{{Schneider} \&
  {Seitz}}{1995}]{Schneider1995}
{Schneider} P.,  {Seitz} C.,  1995, \aap, 294, 411

\bibitem[\protect\citeauthoryear{{Schneider}, {King}  \& {Erben}}{{Schneider}
  et~al.}{2000}]{Schneider2000}
{Schneider} P.,  {King} L.,   {Erben} T.,  2000, \aap, 353, 41

\bibitem[\protect\citeauthoryear{{Shu}, {Zhou}, {Bartelmann}, {Comerford},
  {Huang}  \& {Mellier}}{{Shu} et~al.}{2008}]{Shu2008}
{Shu} C.,  {Zhou} B.,  {Bartelmann} M.,  {Comerford} J.~M.,  {Huang} J.-S.,
  {Mellier} Y.,  2008, \mn@doi [\apj] {10.1086/590049}, 685, 70

\bibitem[\protect\citeauthoryear{{Taylor}, {Dye}, {Broadhurst}, {Benitez}  \&
  {van Kampen}}{{Taylor} et~al.}{1998}]{Taylor1998}
{Taylor} A.~N.,  {Dye} S.,  {Broadhurst} T.~J.,  {Benitez} N.,   {van Kampen}
  E.,  1998, \mn@doi [\apj] {10.1086/305827}, 501, 539

\bibitem[\protect\citeauthoryear{{Umetsu}, {Broadhurst}, {Zitrin}, {Medezinski}
   \& {Hsu}}{{Umetsu} et~al.}{2011}]{Umetsu2011}
{Umetsu} K.,  {Broadhurst} T.,  {Zitrin} A.,  {Medezinski} E.,   {Hsu} L.-Y.,
  2011, \mn@doi [\apj] {10.1088/0004-637X/729/2/127}, \href
  {http://adsabs.harvard.edu/abs/2011ApJ...729..127U} {729, 127}

\bibitem[\protect\citeauthoryear{{Van Waerbeke}, {Hildebrandt}, {Ford}  \&
  {Milkeraitis}}{{Van Waerbeke} et~al.}{2010}]{VanWaerbeke2010}
{Van Waerbeke} L.,  {Hildebrandt} H.,  {Ford} J.,   {Milkeraitis} M.,  2010,
  \mn@doi [\apjl] {10.1088/2041-8205/723/1/L13}, \href
  {http://adsabs.harvard.edu/abs/2010ApJ...723L..13V} {723, L13}

\bibitem[\protect\citeauthoryear{{Wright}}{{Wright}}{2006}]{Wright2006}
{Wright} E.~L.,  2006, \mn@doi [\pasp] {10.1086/510102}, \href
  {http://adsabs.harvard.edu/abs/2006PASP..118.1711W} {118, 1711}

\bibitem[\protect\citeauthoryear{{Wright} \& {Brainerd}}{{Wright} \&
  {Brainerd}}{2000}]{Wright2000}
{Wright} C.~O.,  {Brainerd} T.~G.,  2000, \mn@doi [\apj] {10.1086/308744}, 534,
  34

\makeatother
\end{thebibliography}

\appendix
\onecolumn

\section{Derivation of the mass and multipole moment equivalences for convergence, shear, and flexion}
\subsection{Moments on the full field of view}\label{Appendix:Non-reduced moments of the full field of view}
We compute the general $n$-th order moments within an aperture defined by the weighting function $w(x)$. This allows the application to finite
field observational data. The moments of the mass distribution in terms of $\kappa$ are defined as (SB97)   

\begin{eqnarray}
M^{(n)}_{\kappa}\,=\,\int_0^{\infty}\, dx\,\, x^{n+1}\, w(x)\, \int_0^{2\pi}\, d\varphi\,\, \kappa(x,\varphi, x_0, \varphi_0) \label{Def_M_kappa}
\end{eqnarray}
and the corresponding multipole moments are defined as
\begin{eqnarray}
Q^{(n)}_{\kappa}\,=\,\int_0^{\infty}\, dx\,\, x^{n+1}\,w(x)\,\int_0^{2\pi}\,d\varphi\,\, e^{i n \varphi}\,\kappa(x,\varphi, x_0, \varphi_0). \label{Def_Q_kappa}
\end{eqnarray} 
From now on we will no longer write the explicit dependence on $x_0$ and $\varphi_0$ to simplify the notation. As in (SB97), we can integrate equation (\ref{Def_M_kappa}) by parts to obtain
\begin{eqnarray}
M^{(n)}_{\kappa}\,=\, -\int_0^{\infty}\, dx\,\, x\,W(x)\,\int_0^{2\pi}\,d\varphi\,\,\frac{\partial \kappa(x,\varphi)}{\partial x}, \label{new_M_kappa_part_integration_x}
\end{eqnarray}
where 
\begin{align}
x\, W(x) \,=\, \int\limits_{0}^{x} \, dy\,\, y^{n+1}\, w(y)
\end{align} 
and we require that the boundary term $x\, W(x)\, \kappa(x,\varphi)$ vanishes for $x \rightarrow 0$ and $x \rightarrow \infty$. This can be achieved by picking a weighting function which drops off sufficiently fast. Expressing the radial derivative in Cartesian coordinates and using $\mathcal{F} \,=\, \nabla \kappa$ \citep{Bacon2006}, we can write the mass moments in terms of first flexion:
\begin{eqnarray}
M^{(n)}_{\mathcal{F}}\,=\, \int_0^{\infty}\, dx\,\, x\,W(x)\,\int_0^{2\pi}\,d\varphi\,\, \mathcal{F}_{t}(x,\varphi),
\end{eqnarray}
where 
\begin{eqnarray}
\mathcal{F}_{t}(x,\varphi) \,=\, - \big[\mathcal{F}_{1}(x,\varphi)\, \cos(\varphi) + \mathcal{F}_{2}(x,\varphi)\, \sin(\varphi) \big],\\
F_t(x,\varphi)\,=\,-\big[F_1(x,\varphi) \cos(\varphi) + F_2(x,\varphi) \sin(\varphi) \big],\label{red_tangential_first_F}
\end{eqnarray}
defines the non-reduced and reduced tangential first flexion. We use a mass-sheet reconstruction $\kappa_{\rm rec}$ to find
\begin{eqnarray}
M^{(n)}_{\mathcal{F}}\,=\, \int_0^{\infty}\, dx\,\, x\,W(x)\,\int_0^{2\pi}\,d\varphi\,\, (1-\kappa_{\rm rec}(x,\varphi))\,F_{t}(x,\varphi).
\end{eqnarray}
Note that $\kappa_{\rm rec}$ is different from the convergence that is used to calculate the moments in equations~(\ref{Def_M_kappa}) and (\ref{Def_Q_kappa}). 
Applying the same operations to equation~(\ref{Def_Q_kappa}), we can also express the multipole moments using first flexion:
\begin{eqnarray}
Q^{(n)}_{\mathcal{F}}\,=\, \int_0^{\infty}\, dx\,\, x\,W(x)\,\int_0^{2\pi}\,d\varphi\,\, e^{i n \varphi}\, \mathcal{F}_{t}(x,\varphi)\\
\,=\, \int_0^{\infty}\, dx\,\, x\,W(x)\,\int_0^{2\pi}\,d\varphi\,\, e^{i n \varphi}\, (1-\kappa_{\rm rec}(x,\varphi))\, F_{t}(x,\varphi).
\end{eqnarray}\\

The moments can also be expressed using a combination of $\mathcal{F}$ and $\mathcal{G}$. We define the tangential and radial shear and the corresponding reduced quantities by
\begin{align}
\gamma_{t}(x,\varphi) \,=\, -\big[\gamma_{1}(x,\varphi)\, \cos(2\varphi) + \gamma_{2}(x,\varphi)\, \sin(2\varphi) \big],  \\
\gamma_{r}(x,\varphi) \,=\, -\big[\gamma_{2}(x,\varphi)\, \cos(2\varphi) - \gamma_{1}(x,\varphi)\, \sin(2\varphi) \big], \\
g_{t}(x,\varphi) \,=\, -\big[g_{1}(x,\varphi)\, \cos(2\varphi) + g_{2}(x,\varphi)\, \sin(2\varphi) \big], \\
g_{r}(x,\varphi) \,=\, -\big[g_{2}(x,\varphi)\, \cos(2\varphi) - g_{1}(x,\varphi)\, \sin(2\varphi) \big],
\end{align}
and use the relation shown in \citet{Kaiser1995a}
\begin{eqnarray}
\nabla \kappa \,=\, 
\left(\begin{array}{c}
\gamma_{1,1} + \gamma_{2,2} \\
\gamma_{2,1} - \gamma_{1,2}
\end{array}\right). \label{Rel_nabla_kappa_shear}
\end{eqnarray}
Inserting the radial part of this relation transformed to polar coordinates into equation~(\ref{new_M_kappa_part_integration_x}), using trigonometric identities, integrating by parts and demanding that the surface terms $x W(x) \gamma_t (x, \varphi)$ vanish for $x \to 0$, $\infty$ leads to an expression of the mass moments using shear (SB97),
\begin{align}
M^{(n)}_{\gamma} \,=\, \int\limits_{0}^{\infty}\, dx\,\, \big[ 2 W(x) - x^{n+1}\, w(x) \big] \int\limits_{0}^{2 \pi}\, d\varphi\,\, \gamma_{t}(x,\varphi)   
\,=\, \int\limits_{0}^{\infty}\, dx\,\, \big[ 2 W(x) - x^{n+1}\, w(x) \big] \int\limits_{0}^{2 \pi}\, d\varphi\,\, (1-\kappa_{\rm rec}(x,\varphi)) g_{t}(x,\varphi). \label{Def_mass_mom_shear}
\end{align}
We define
\begin{eqnarray}
V(x) \,=\, \int\limits_{0}^{x}\, dy\,\, W(y)
\end{eqnarray}
and require that $\big[2 V(x) - x\,W(x) \big]\, \gamma_{t}(x,\varphi) \rightarrow 0  \,\,\text{for}\, x \rightarrow 0,\, x \rightarrow \infty$.
Then the boundary terms resulting from a partial integration with respect to $x$ vanish and we find
\begin{align}
M^{(n)}_{\gamma} \,=\, -\int_0^{\infty}\, dx\,\, \big[2 V(x)\, - x\, W(x) \big] \int_0^{2\pi}\,d\varphi\,\, 
\frac{\partial \gamma_{t}(x,\varphi)}{\partial x}.
\end{align}
We further define the radial first flexion, the tangential and radial second flexion and their reduced counterparts by
\begin{align}
\mathcal{F}_{r}(x,\varphi) \,=\, -\big[ \mathcal{F}_{2}(x,\varphi)\, \cos(\varphi) - \mathcal{F}_{1}(x,\varphi)\, \sin(\varphi) \big],  \\
\mathcal{G}_{t}(x,\varphi) \,=\, -\big[ \mathcal{G}_{1}(x,\varphi)\, \cos(3 \varphi) + \mathcal{G}_{2}(x,\varphi)\, \sin(3 \varphi) \big], \\
\mathcal{G}_{r}(x,\varphi) \,=\, -\big[ \mathcal{G}_{2}(x,\varphi)\, \cos(3 \varphi) - \mathcal{G}_{1}(x,\varphi)\, \sin(3 \varphi) \big],\\
F_{r}(x,\varphi) \,=\, -\big[ F_{2}(x,\varphi)\, \cos(\varphi) - F_{1}(x,\varphi)\, \sin(\varphi) \big],\\
G_{t}(x,\varphi) \,=\, -\big[ G_{1}(x,\varphi)\, \cos(3 \varphi) + G_{2}(x,\varphi)\, \sin(3 \varphi) \big], \\
G_{r}(x,\varphi) \,=\, -\big[ G_{2}(x,\varphi)\, \cos(3 \varphi) - G_{1}(x,\varphi)\, \sin(3 \varphi) \big].
\end{align}
Now we can express the radial derivative in Cartesian coordinates, apply trigonometric identities and use the relations between flexion and shear derivatives in equation~(\ref{Def_flexion_shear_deriv})
to obtain the mass moments in terms of first and second flexion:
\begin{align}
M^{(n)}_{\mathcal{F},\mathcal{G}} \,=\, -\int_0^{\infty}\, dx\,\, \big[V(x)\, - \frac{1}{2} x\, W(x) \big]  \int_0^{2\pi}\,d\varphi\,\, \big[ \mathcal{F}_{t}(x,\varphi) + \mathcal{G}_{t}(x,\varphi) \big]  \\
\,=\, -\int_0^{\infty}\, dx\,\, \big[V(x)\, - \frac{1}{2} x\, W(x) \big]\, \int_0^{2\pi}\,d\varphi\,\,  \big[ (1-\kappa_{\rm rec}(x,\varphi)) F_{t}(x,\varphi) + (1-\kappa_{\rm rec}(x,\varphi)) G_{t}(x,\varphi) \big].
\end{align}\\
The multipole moments using $\mathcal{F}$ and $\mathcal{G}$ can be derived in an analogous way. We integrate equation~(\ref{Def_Q_kappa}) by parts and require that the boundary term $x\, W(x)\, \kappa(x,\varphi)$ vanishes for $x \rightarrow 0$ and $x \rightarrow \infty$. Inserting the radial part of equation~(\ref{Rel_nabla_kappa_shear}) transformed to polar coordinates,  integrating by parts and demanding that $x W(x) \gamma_t(x,\varphi)$ vanishes for $x \to 0$, $\infty$ leads to the multipole moments in terms of shear (SB97),
\begin{align}
Q^{(n)}_{\gamma} \,=\, \int_0^{\infty}\, dx\,\, \big[2 W(x)\, - x^{n+1}\, w(x)\big] \int_0^{2\pi}\,d\varphi\,\, e^{i n \varphi}\, \gamma_{t}(x,\varphi)  - i\, n\, \int\limits_{0}^{\infty}\, dx\,\, W(x)\, \int\limits_{0}^{2 \pi}\, d\varphi\,\, e^{i n \varphi}\, \gamma_{r}(x,\varphi)\\
\,=\, \int_0^{\infty}\, dx\,\, \big[2 W(x)\, - x^{n+1}\, w(x)\big] \int_0^{2\pi}\,d\varphi\,\, e^{i n \varphi}\, (1-\kappa_{\rm rec}(x,\varphi)) g_{t}(x,\varphi)  - i\, n\, \int\limits_{0}^{\infty}\, dx\,\, W(x)\, \int\limits_{0}^{2 \pi}\, d\varphi\,\, e^{i n \varphi}\, (1-\kappa_{\rm rec}(x,\varphi)) g_{r}(x,\varphi).
\end{align}

Partial integration with respect to $x$ and applying the same transformations and identities as before leads to 
\begin{align}
Q^{(n)}_{\mathcal{F},\mathcal{G}} \,=\, -\int_0^{\infty}\, dx\,\, \big[V(x)\, - \frac{1}{2} x\, W(x) \big] \int_0^{2\pi}\,d\varphi\,\, e^{i n \varphi}\,  \big[ \mathcal{F}_{t}(x,\varphi) + \mathcal{G}_{t}(x,\varphi) \big] \nonumber \\
+ i\, \frac{n}{2}\, \int\limits_{0}^{\infty}\, dx\,\, V(x)\, \int\limits_{0}^{2 \pi}\, d\varphi\,\, e^{i n \varphi}\, 
\big[ \mathcal{F}_{r}(x,\varphi) + \mathcal{G}_{r}(x,\varphi) \big]\\
\,=\,-\int_0^{\infty}\, dx\,\, \big[V(x)\, - \frac{1}{2} x\, W(x) \big] \int_0^{2\pi}\,d\varphi\,\, e^{i n \varphi}\,  \big[ (1-\kappa_{\rm rec}(x,\varphi)) F_{t}(x,\varphi) + (1-\kappa_{\rm rec}(x,\varphi)) G_{t}(x,\varphi) \big] \nonumber \\
+ i\, \frac{n}{2}\, \int\limits_{0}^{\infty}\, dx\,\, V(x)\, \int\limits_{0}^{2 \pi}\, d\varphi\,\, e^{i n \varphi}\,  \big[ (1-\kappa_{\rm rec}(x,\varphi)) F_{r}(x,\varphi) + (1-\kappa_{\rm rec}(x,\varphi)) G_{r}(x,\varphi) \big],
\end{align}
where we required that
\begin{align} 
\big[2 V(x) - x\,W(x) \big]\, \gamma_{t}(x,\varphi) \rightarrow 0 \,\, \text{for}\, x \rightarrow 0,\, x \rightarrow \infty, \label{Req_mass_mom_F_G}\\
V(x)\, \gamma_{r}(x,\varphi) \rightarrow 0  \,\,\text{for}\, x \rightarrow 0,\, x \rightarrow \infty,
\end{align}
so that the boundary terms in the partial integration vanish.

\subsubsection{Moments on rings}\label{Appendix: Non-reduced moments on rings}
We now extend our moment formulas to rings, i.e. we omit the innermost part of the lens in the integration, because we typically do not have weak lensing information in this area. The derivation is exactly analogous to the one presented in the previous section and we will therefore only state the resulting equivalent formulas and the respective requirements for the weighting function. Furthermore we will not explicitly denote the reconstruction of shear and flexion using $\kappa_{rec}$ to keep this section brief. We define
\begin{eqnarray}
M^{(n)}_{R; \kappa}\,=\,\int_R^{\infty}\, dx\,\, x^{n+1}\, w(x)\, \int_0^{2\pi}\, d\varphi\,\, \kappa(x,\varphi)
\end{eqnarray}
and the corresponding multipole moments
\begin{eqnarray}
Q^{(n)}_{R; \kappa}\,=\,\int_R^{\infty}\, dx\,\, x^{n+1}\,w(x)\,\int_0^{2\pi}\,d\varphi\,\, e^{i n \varphi}\,\kappa(x,\varphi).
\end{eqnarray}  
We further define
\begin{eqnarray}
x W_R(x)\,=\, \int_{R}^{x}~dy~y^{n+1} w(y)
\end{eqnarray}
and demand that $x W_R(x) \kappa(x,\varphi)$ vanishes for $x\to R,\infty$. Note that this is simply a different choice of the integration constant. We obtain
\begin{eqnarray}
M^{(n)}_{R;\mathcal{F}}\,=\, \int_R^{\infty}\, dx\,\, x\,W_R(x)\,\int_0^{2\pi}\,d\varphi\,\,\mathcal{F}_{t}(x,\varphi),\\
Q^{(n)}_{R;\mathcal{F}}\,=\, \int_R^{\infty}\, dx\,\, x\,W_R(x)\,\int_0^{2\pi}\,d\varphi\,\, e^{i n \varphi}\, \mathcal{F}_{t}(x,\varphi).
\end{eqnarray}
We further demand that the term $x W_R(x) \gamma_t(x,\varphi)$ vanishes for $x\to R,\infty$ and note that the other surface terms cancel due to periodicity. We have now
\begin{align}
M^{(n)}_{R;\gamma} \,=\, \int\limits_{R}^{\infty}\, dx\,\, \big[ 2 W_R(x) - x^{n+1}\, w(x) \big] \int\limits_{0}^{2 \pi}\, d\varphi\,\, \gamma_{t}(x,\varphi),\\
Q^{(n)}_{R;\gamma} \,=\, \int_R^{\infty}\, dx\,\, \big[2 W_R(x)\, - x^{n+1}\, w(x)\big] \int_0^{2\pi}\,d\varphi\,\, e^{i n \varphi}\, \gamma_{t}(x,\varphi) - i\, n\, \int\limits_{R}^{\infty}\, dx\,\, W_R(x)\, \int\limits_{0}^{2 \pi}\, d\varphi\,\, e^{i n \varphi}\, \gamma_{r}(x,\varphi).
\end{align}
We define
\begin{eqnarray}
V_R(x)\,=\, \int_{R}^{x}~dy~ W_R(y)
\end{eqnarray}
and require that $[2V_R(x) - x W_R(x)] \gamma_t(x,\varphi)$ and, for the multipole moments, $V_R(x) \gamma_r(x,\varphi)$ vanish for $x\to R,\infty$ to find
\begin{eqnarray}
M^{(n)}_{R;\mathcal{F},\mathcal{G}} \,=\, -\int_R^{\infty}\, dx\,\, \big[V_R(x)\, - \frac{1}{2} x\, W_R(x) \big]  \int_0^{2\pi}\,d\varphi\,\, \big[ \mathcal{F}_{t}(x,\varphi) + \mathcal{G}_{t}(x,\varphi) \big],\\
Q^{(n)}_{R;\mathcal{F},\mathcal{G}} \,=\,
-\int_R^{\infty}\, dx\,\, \big[V_R(x)\, - \frac{1}{2} x\, W_R(x) \big] \int_0^{2\pi}\,d\varphi\,\, e^{i n \varphi}\, \big[ \mathcal{F}_{t}(x,\varphi) + \mathcal{G}_{t}(x,\varphi) \big] \nonumber \\
+ i\, \frac{n}{2}\, \int\limits_{R}^{\infty}\, dx\,\, V_R(x)\, \int\limits_{0}^{2 \pi}\, d\varphi\,\, e^{i n \varphi}\, 
\big[ \mathcal{F}_{r}(x,\varphi) + \mathcal{G}_{r}(x,\varphi) \big].
\end{eqnarray}
\\
If weak lensing data is only available in a small part of the ring, we can further restrict the integration to a partial ring. For this purpose, we modify the mass and multipole moment equations for $\kappa$ and $\mathcal{F}$ by replacing $2\pi$ with the desired maximum angle $\phi$ and their equivalence still holds. However, they are now no longer equivalent to the $\gamma$ and $\mathcal{F},\mathcal{G}$ moments. The integration area of these moments cannot be restricted in general, since the derivation of the equivalences uses the periodicity due to the $2\pi$ boundary. However, in the case of the mass moments $M^{(n)}_{R;\gamma}$ and $M^{(n)}_{R;\mathcal{F},\mathcal{G}}$, replacing $2\pi$ with $\pi$ also preserves the required periodicity and thus it is possible to restrict these moments to one half of a ring. Therefore all mass moments are also equivalent for $\phi \,=\, \pi$.

\section{Derivation of the mass and multipole moment equivalences for $K$, reduced shear, and reduced flexions}\label{Appendix: Reduced moments on rings}

We derive the moment equivalence relations for the reduced quantities. Following \citet{Cain2011}, we define 
\begin{eqnarray}
K(x,\varphi,x_0, \varphi_0) \,=\, -\ln(1-\kappa(x,\varphi,x_0,\varphi_0))
\end{eqnarray}
and thus
\begin{eqnarray}
K_{,i}\,=\,\frac{1}{1-\kappa} \kappa_{,i}\,=\,F_{i}.\label{red_K_red_F_relation}
\end{eqnarray}
We want $\kappa < 1$ and choose the lower integral limit for our ring, $R$, such that $\kappa(x,\varphi) < 1~ \forall x \geq R,\forall \varphi$. Therefore $K(x,\varphi)$ is well-defined and finite and we can define
\begin{eqnarray}
M^{(n)}_K \,=\, \int_R^{\infty}\,dx\,x^{n+1}\,w(x) \int_{0}^{2\pi}\,d\varphi\, K(x,\varphi,x_0,\varphi_0),\label{red_M_K}\\
Q^{(n)}_K \,=\, \int_R^{\infty}\,dx\,x^{n+1}\,w(x) \int_{0}^{2\pi}\,d\varphi\, e^{i n \varphi}\, K(x,\varphi,x_0,\varphi_0).
\end{eqnarray}
In the following, we will again not explicitly denote $x_0$ and $\varphi_0$ to keep the notation simple. We require that $x W_R(x) K(x,\varphi)$ vanishes for $x\to R$ and $x\to \infty$ and integrate equation~(\ref{red_M_K}) by parts with respect to $x$ to find 
\begin{eqnarray}
M^{(n)}_K\,=\,-\int_R^{\infty}\,dx\,x W_R(x) \int_{0}^{2\pi}\,d\varphi\, \frac{\partial K(x,\varphi)}{\partial x}.\label{red_F_K_partial_derivative}
\end{eqnarray}
We now express the radial derivative in Cartesian coordinates, apply equation~(\ref{red_K_red_F_relation}) and use the defininition of $F_t$ in equation~(\ref{red_tangential_first_F}). Then we have
\begin{eqnarray}
M^{(n)}_F\,=\,\int_R^{\infty}\,dx\,x W_R(x) \int_{0}^{2\pi}\,d\varphi\, F_t(x,\varphi).
\end{eqnarray}
In the exact same way we find for the multipole moments
\begin{eqnarray}
Q^{(n)}_F\,=\,\int_R^{\infty}\,dx\,x W_R(x) \int_{0}^{2\pi}\,d\varphi\, e^{i n \varphi}\, F_t(x,\varphi).
\end{eqnarray}
Now we would like to express the moments using the reduced shear. Therefore we rewrite the $K$ derivatives by using equation~(\ref{Rel_nabla_kappa_shear}) and the definition of $g$:
\begin{eqnarray}
K_{,1} \,=\, g_{1,1} - g_1 K_{,1} + g_{2,2} - g_2 K_{,2}\label{red_shear_1},\\
K_{,2} \,=\, g_{2,1} - g_2 K_{,1} - g_{1,2} + g_1 K_{,2}\label{red_shear_2}.
\end{eqnarray}
Furthermore, we use $K_{,i} = F_i$ to find
\begin{eqnarray}
\frac{\partial K}{\partial x}\,=\, -F_t,\\
\frac{\partial K}{\partial \varphi}\,=\, -x F_r.
\end{eqnarray}
Transforming the Cartesian derivatives on the right hand side of equations~(\ref{red_shear_1}) and (\ref{red_shear_2}) into polar coordinates, using trigonometric identities and the definitions of the tangential and radial reduced shear and reduced flexion, we have
\begin{eqnarray}
\frac{\partial K}{\partial x}\,=\, \cos(\varphi)\, K_{,1} + \sin(\varphi)\, K_{,2} = -\big[\frac{\partial g_t}{\partial x} - \frac{\cos(2\varphi)}{x}\frac{\partial g_2}{\partial \varphi} + \frac{\sin(2\varphi)}{x}\frac{\partial g_1}{\partial \varphi} + g_t F_t + g_r F_r \big].
\end{eqnarray}
We insert this expression into equation~(\ref{red_F_K_partial_derivative}), integrate the term with the $x$-derivative by parts and demand that $x W_R(x) g_t(x,\varphi)$ vanishes for $x\to R$ and $x\to \infty$. Then the surface term vanishes and we have
\begin{eqnarray}
M^{(n)}_F \,=\, \int_{R}^{\infty}\,dx\, x W_R(x) \int_{0}^{2\pi}\,d\varphi\, (g_t F_t + g_r F_r) - \int_{R}^{\infty}\,dx\,x^{n+1} w(x) \int_{0}^{2\pi}\, d\varphi\, g_t \nonumber\\
- \int_{R}^{\infty}\, dx\, W_R(x) \int_{0}^{2\pi}\, d\varphi\, (\cos(2\varphi)\frac{\partial g_2}{\partial \varphi} - \sin(2\varphi)\frac{\partial g_1}{\partial \varphi}).
\end{eqnarray} 
Integrating the $\varphi$ derivatives by parts and noting that the surface terms vanish due to their periodicity, we find
\begin{eqnarray}
M^{(n)}_g &\,=\,& \int_{R}^{\infty} dx\, x W_R(x) \int_{0}^{2\pi}\,d\varphi\, (g_t(x,\varphi) F_t (x,\varphi) + g_r(x,\varphi) F_r(x,\varphi)) \nonumber\\
&&+ \int_R^{\infty} dx\, (2 W_R(x) - x^{n+1} w(x)) \int_{0}^{2\pi} d\varphi\, g_t(x,\varphi).\label{red_M_shear}
\end{eqnarray}
Using the same steps, we can derive the multipole moments in terms of reduced shear and reduced flexion,
\begin{eqnarray}
Q^{(n)}_g &\,=\,& \int_{R}^{\infty} dx\, x W_R(x) \int_{0}^{2\pi}\,d\varphi\, e^{in\varphi} (g_t(x,\varphi) F_t (x,\varphi) + g_r(x,\varphi) F_r(x,\varphi)) \nonumber\\
&&+ \int_R^{\infty} dx\, (2 W_R(x) - x^{n+1} w(x)) \int_{0}^{2\pi} d\varphi\,e^{in\varphi}\, g_t(x,\varphi) -i n \int_{R}^{\infty}\,dx\, W_R(x) \int_{0}^{2\pi}\, d\varphi\, e^{in\varphi} g_r(x,\varphi).\label{red_Q_shear}
\end{eqnarray}
We can integrate the second term in equation~(\ref{red_M_shear}) by parts with respect to $x$ and demand that the surface term $(2 V_R(x) - x W_R(x)) g_t(x,\varphi)$ vanishes for $x\to R$ and $x\to \infty$. We have now
\begin{eqnarray}
M^{(n)}_g &\,=\,& \int_{R}^{\infty} dx\, x W_R(x) \int_{0}^{2\pi}\,d\varphi\, (g_t(x,\varphi) F_t (x,\varphi) + g_r(x,\varphi) F_r(x,\varphi)) \nonumber \\
&&  - \int_R^{\infty} dx\, (2 V_R(x) - x W_R(x)) \int_{0}^{2\pi} d\varphi\, \frac{\partial g_t(x,\varphi)}{\partial x}.
\end{eqnarray}
Again applying a transformation to Cartesian coordinates, trigonometric identities, $\nabla \kappa = \mathcal{F}$, and the relations from equation~(\ref{Def_flexion_shear_deriv}), we find for the derivative
\begin{eqnarray}
\frac{\partial g_t}{\partial x} \,=\, \frac{1}{2}(F_t + G_t) - g_t F_t
\end{eqnarray}
and thus we have for the mass moment
\begin{eqnarray}
M^{(n)}_{F,G} &\,=\,& \int_{R}^{\infty} dx\, x W_R(x) \int_{0}^{2\pi}\,d\varphi\,  (g_r(x,\varphi) F_r(x,\varphi) + \frac{1}{2}(F_t(x,\varphi) + G_t(x,\varphi))) \nonumber\\
&&- \int_R^{\infty} dx\, V_R(x) \int_{0}^{2\pi} d\varphi\,  (F_t(x,\varphi) + G_t(x,\varphi) - 2 g_t(x,\varphi) F_t(x,\varphi)).
\end{eqnarray}
Analogously, we can integrate the last two terms in equation~(\ref{red_Q_shear}) by parts and demand that the resulting surface terms $(2V_R(x) - xW_R(x)) g_t(x,\varphi)$ and $V_R(x) g_r(x,\varphi)$ vanish for $x\to R$ and $x\to \infty$. In the same way as above, we can calculate the derivative as
\begin{eqnarray}
\frac{\partial g_r}{\partial x} \,=\, \frac{1}{2}(F_r + G_r) -g_r F_t
\end{eqnarray}
and use our result for $\partial g_t / \partial x$ to find
\begin{eqnarray}
Q^{(n)}_{F,G} &\,=\,& \int_{R}^{\infty} dx\, x W_R(x) \int_{0}^{2\pi}\,d\varphi\,e^{in\varphi} (g_t(x,\varphi) F_t(x,\varphi) + g_r(x,\varphi) F_r (x,\varphi)) - \int_{R}^{\infty}\,dx\,(2V_R(x) - xW_R(x)) \int_{0}^{2\pi}\,d\varphi\, e^{in\varphi} \nonumber\\
&& \cdot (\frac{1}{2}(F_t(x,\varphi) + G_t(x,\varphi)) - g_t(x,\varphi) F_t(x,\varphi))  + in \int_R^{\infty} dx\, V_R(x) \int_{0}^{2\pi} d\varphi\, e^{in\varphi} \nonumber\\
&& \cdot (\frac{1}{2}(F_r(x,\varphi) + G_r(x,\varphi)) - g_r(x,\varphi) F_t(x,\varphi)).
\end{eqnarray}
\\
If weak lensing data is only available in a small angular window, we can restrict the integration area to a partial ring. We modify the mass and multipole moment equations for $K$ and $F$ by replacing $2\pi$ with the maximum angle $\phi$ and they will still be equivalent. However, they are then no longer equivalent to the $g$ and $F,G$ moments. The integration area of these moments cannot be further restricted, because otherwise the cancellation of the surface terms due to the periodicity would no longer hold.

\section{Mass moment dispersions}\label{Appendix:Moment dispersions}

\subsection{Shear, flexion, and convergence moments}
We calculate the mass moment dispersions in the absence of lensing. For the convergence, we have $< M_{\kappa}^{(n)} > = 0$ and
\begin{align}
\sigma_{M,\kappa}^2 = <|M_{\kappa}^{(n)}|^2>\\
= \int\limits_{0}^{\infty}\int\limits_{0}^{\infty}~dx~dy~x^{n+1} y^{n+1} w(x) w(y) \int\limits_{0}^{2\pi}\int\limits_{0}^{2\pi}~d\varphi~d\varphi' <\kappa(x,\varphi) \kappa(y,\varphi')>.
\end{align}
Following \citet{Schneider2000}, the magnification signal for number counts is $|\Delta n_{\kappa}| = |\mu^{\beta - 1} - 1| n_{\kappa}$, where $\mu$ is the magnification and $n_{\kappa}$ is the source density. The Poisson noise is $\sqrt{N_{\kappa, a}}$, where $N_{\kappa, a}$ is the number of sources over which we average to determine $\kappa$ in a given area $a$. We neglect the additional noise due to source clustering. In the weak lensing regime and assuming $\beta = 0.5$, we have  $\mu \approx 1 + 2\kappa$ and thus $|\kappa| \approx |\Delta n_{\kappa}| / n_{\kappa}$.
\\
\\
We assume a discrete distribution of $\kappa$ over an area $A$ and measure the convergence by averaging our source counts in an area $a$. Therefore our measurement points of $\kappa$ are uncorrelated. Thus we have for the sum of the convergence variance over the area A:
\begin{align}
\frac{n_{\kappa}^2 a^2}{n_{\kappa}^2} \sum\limits_{i=1}^{N_{\kappa}/n_{\kappa}a} \sum\limits_{j=1}^{N_{\kappa}/n_{\kappa}a} < \kappa_i \kappa_j> = a^2 \sum\limits_{i=1}^{N_{\kappa}/n_{\kappa}a} \frac{1}{n_{\kappa} a} = \frac{A}{n_{\kappa}},\\
<\kappa_i \kappa_j> = \frac{1}{n_{\kappa} a} \delta_{ij}.
\end{align}
Extending this to integration, we have
\begin{align}
\int \int~dx^2 dy^2~<\kappa(\vec{x}) \kappa(\vec{y})> = \frac{1}{n_{\kappa}} \int dx^2 = \frac{A}{n_{\kappa}}
\end{align}
and thus
\begin{align}
<\kappa(\vec{x}) \kappa(\vec{y})> = \frac{1}{n_{\kappa}} \delta(\vec{x} - \vec{y})
\end{align}
with the Dirac delta distribution $\delta(\vec{x}-\vec{y})$.
\\
\\
Therefore we have 
\begin{align}
\sigma_{M,\kappa}^2 = \frac{2\pi}{n_{\kappa}} \int\limits_{0}^{\infty}~dx~x^{2n + 1} w(x)^2
\end{align} 
and thus the standard deviation
\begin{align}
\sigma_{M,\kappa} = \sqrt{\frac{2\pi}{n_{\kappa}}} \Bigg( \int\limits_{0}^{\infty}~dx~x^{2n+1} w(x)^2 \Bigg)^{\frac{1}{2}}.
\end{align}
\\
\\
Before calculating the shear and flexion moments, we make the following observation. In the case of a discrete distribution over an area $A$ and no lensing, all our measurements of the shear are uncorrelated and we have
\begin{align}
\frac{1}{n_{\gamma}^2} \sum\limits_{i=1}^N \sum\limits_{j=1}^N < \gamma_i \gamma_j> = \frac{1}{n_{\gamma}^2} \sum\limits_{i=1}^{N} \sigma_{\gamma}^2 = \frac{A}{n_{\gamma}} \sigma_{\gamma}^2,\\
<\gamma_i \gamma_j> = \sigma_{\gamma}^2 \delta_{ij}
\end{align}
with the number density $n_{\gamma} = N_{\gamma}/A$, where $N_{\gamma}$ is the number of sources where we measured $\gamma$. Extending this to integration, we have
\begin{align}
\int \int~dx^2 dy^2~<\gamma(\vec{x}) \gamma(\vec{y})> = \frac{1}{n_{\gamma}} \int dx^2~\sigma_{\gamma}^2 = \frac{A}{n_{\gamma}} \sigma_{\gamma}^2
\end{align}
and thus
\begin{align}
<\gamma(\vec{x}) \gamma(\vec{y})> = \frac{\sigma_{\gamma}^2}{n_{\gamma}} \delta(\vec{x} - \vec{y})
\end{align}
with the Dirac delta distribution $\delta(\vec{x}-\vec{y})$.
\\
\\
The errors on $\kappa_{rec}$ and $g_t$ generally depend on the measurement technique. However, we are assuming the absence of a lens for the calculation of the standard deviation, so we can look at the mass moment in terms of $\gamma$ instead of $\kappa_{rec}$ and $g_t$ and we neglect the measurement noise. Therefore we have $<M_{\gamma}^{(n)}> = 0$ and for the variance:
\begin{align}
\sigma_{M,\gamma}^2 = <|M_{\gamma}^{(n)}|^2>\\
= \int\limits_{0}^{\infty}\int\limits_{0}^{\infty}~dx~dy~[2W(x) - x^{n+1} w(x)] [2W(y) - y^{n+1} w(y)] \int\limits_{0}^{2\pi}\int\limits_{0}^{2\pi}~d\varphi~d\varphi'~<\gamma_t(x,\varphi) \gamma_t(y,\varphi')>.
\end{align}
We use
\begin{align}
<\gamma_t(x,\varphi) \gamma_t(y,\varphi')> = \frac{\sigma_{\gamma}^2}{n_{\gamma}} \frac{1}{y} \delta(x-y) \delta(\varphi - \varphi'),
\end{align}
where we have written the Dirac delta distribution in polar coordinates, and get
\begin{align}
\sigma_{M,\gamma}^2 = \int\limits_{0}^{\infty}~dx~\frac{1}{x} [2W(x) - x^{n+1} w(x)]^2 \int\limits_{0}^{2\pi}~d\varphi~\frac{\sigma_{\gamma}^2}{n_{\gamma}}
\end{align}
and for the standard deviation
\begin{align}
\sigma_{M,\gamma} = \sqrt{\frac{2\pi}{n_{\gamma}}} \sigma_{\gamma} \Bigg(\int\limits_{0}^{\infty}~dx~\frac{1}{x} [2W(x) - x^{n+1} w(x)]^2 \Bigg)^{\frac{1}{2}}.
\end{align}
\\
\\
Similarly, we have $<M_{\mathcal{F}}^{(n)}> = 0$ and
\begin{align}
\sigma_{M,\mathcal{F}}^2 = <|M_{\mathcal{F}}^{(n)}|^2>\\
= \int\limits_{0}^{\infty}\int\limits_{0}^{\infty}~dx~dy~x W(x) y W(y) \int\limits_{0}^{2\pi}\int\limits_{0}^{2\pi}~d\varphi~d\varphi'~<\mathcal{F}_t(x,\varphi) \mathcal{F}_t(y,\varphi')>.
\end{align}
Using
\begin{align}
<\mathcal{F}_t(x,\varphi) \mathcal{F}_t(y,\varphi')> = \frac{\sigma_{\mathcal{F}}^2}{n_{\mathcal{F}}} \frac{1}{y} \delta(x-y) \delta(\varphi - \varphi'), \label{eqn:F_Flexion_Dispersion}
\end{align}
we have
\begin{align}
\sigma_{M,\mathcal{F}}^2 = \int\limits_{0}^{\infty}~dx~x W(x)^2 \int\limits_{0}^{2\pi}~d\varphi~\frac{\sigma_{\mathcal{F}}^2}{n_{\mathcal{F}}}
\end{align}
and consequently
\begin{align}
\sigma_{M,\mathcal{F}} = \sqrt{\frac{2\pi}{n_{\mathcal{F}}}} \sigma_{\mathcal{F}} \Bigg(\int\limits_{0}^{\infty}~dx~xW(x)^2\Bigg)^{\frac{1}{2}}.
\end{align}
\\
\\
In the same way, we get  $<M_{\mathcal{F},\mathcal{G}}^{(n)}> = 0$ and
\begin{align}
\sigma_{M,\mathcal{F},\mathcal{G}}^2 = <|M_{\mathcal{F},\mathcal{G}}^{(n)}|^2>\\
= \int\limits_{0}^{\infty}\int\limits_{0}^{\infty}~dx~dy~[V(x) - \frac{1}{2} x W(x)] [V(y) - \frac{1}{2} y W(y)] \int\limits_{0}^{2\pi}\int\limits_{0}^{2\pi}~d\varphi~d\varphi'~(<\mathcal{F}_t(x,\varphi)\mathcal{F}_t(y,\varphi')> \\\nonumber
+ <\mathcal{F}_t(x,\varphi)\mathcal{G}_t(y,\varphi')> + <\mathcal{G}_t(x,\varphi)\mathcal{F}_t(y,\varphi')> + <\mathcal{G}_t(x,\varphi)\mathcal{G}_t(y,\varphi')>).
\end{align}
We use equation~(\ref{eqn:F_Flexion_Dispersion}) and
\begin{align}
<\mathcal{G}_t(x,\varphi)\mathcal{G}_t(y,\varphi')> = \frac{\sigma_{\mathcal{G}}^2}{n_{\mathcal{G}}} \frac{1}{y} \delta(x-y) \delta(\varphi - \varphi')
\end{align}
and assume that
\begin{align}
<\mathcal{F}_t(x,\varphi)\mathcal{G}_t(y,\varphi')> = 0
\end{align}
to get
\begin{align}
\sigma_{M,\mathcal{F},\mathcal{G}}^2 = \int\limits_{0}^{\infty}~dx~\frac{1}{x} [V(x) - \frac{1}{2} x W(x)]^2 \int\limits_{0}^{2\pi}~d\varphi~(\frac{\sigma_{\mathcal{F}}^2}{n_{\mathcal{F}}} + \frac{\sigma_{\mathcal{G}}^2}{n_{\mathcal{G}}})
\end{align}
and the standard deviation
\begin{align}
\sigma_{M,\mathcal{F},\mathcal{G}} = \sqrt{2\pi} \sqrt{\frac{\sigma_{\mathcal{F}}^2}{n_{\mathcal{F}}} + \frac{\sigma_{\mathcal{G}}^2}{n_{\mathcal{G}}}} \Bigg(\int\limits_{0}^{\infty}~dx~\frac{1}{x} [V(x) - \frac{1}{2} x W(x)]^2 \Bigg)^{\frac{1}{2}}.
\end{align}

\subsection{Reduced shear, reduced flexion, and K moments}
We have 
\begin{align}
< M_K^{(n)} > =  \int\limits_{R}^{\infty}~dx~x^{n+1} w(x) \int\limits_{0}^{2\pi}~d\varphi~< K(x,\varphi) >.
\end{align}
In the absence of lensing, $\kappa$ is typically everywhere smaller than 1, so we can use the Mercator series,
\begin{align}
\ln(1+x) = \sum\limits_{n=1}^{\infty} \frac{(-1)^{n+1}}{n}x^n~\text{for~} |x| < 1,\label{eqn:Mercator_series}
\end{align}
to find
\begin{align}
< K(x,\varphi) > = (-1) \sum\limits_{n=1}^{\infty} \frac{(-1)^{n+1}}{n} (-1)^n < \kappa(x,\varphi)^n >.
\end{align}
As $\kappa$ is small, we can restrict ourselves to the 2 lowest order terms,
\begin{align}
<\kappa(x,\varphi)> = 0,\\
<\kappa(x,\varphi) \kappa(x,\varphi)> = \frac{1}{n_{\kappa} a},
\end{align}
where we again average the sources over the area $a$, and get
\begin{align}
< K(x,\varphi) > = \frac{1}{2 n_{\kappa} a}.
\end{align}
Therefore we have
\begin{align}
< M_K^{(n)} > =  \frac{\pi}{n_{\kappa} a} \int\limits_{R}^{\infty}~dx~x^{n+1} w(x).
\end{align}
The variance is
\begin{align}
\sigma_{M,K}^2 = <|M_{K}^{(n)}|^2> - < M_K^{(n)}>^2.
\end{align}
We have
\begin{align}
<|M_{K}^{(n)}|^2> = \int\limits_{R}^{\infty}~\int\limits_{R}^{\infty}~dx~dy~x^{n+1} y^{n+1} w(x) w(y) \int\limits_{0}^{2\pi} \int\limits_{0}^{2\pi}~d\varphi~d\varphi'~<K(x,\varphi) K(y,\varphi')> 
\end{align}
and
\begin{align}
<K(x,\varphi) K(y,\varphi')> = < \ln(1-\kappa(x,\varphi)) \ln(1-\kappa(y,\varphi'))>
\end{align}
and using again the Mercator series, we get
\begin{align}
<K(x,\varphi) K(y,\varphi')> = < \sum\limits_{n=1}^{\infty} \frac{(-1)^{n+1}}{n}(-1)^{n}\kappa^n(x,\varphi)~\sum\limits_{m=1}^{\infty} \frac{(-1)^{m+1}}{m}(-1)^{m}\kappa^m(y,\varphi') >\\
= < \sum\limits_{n=1}^{\infty} \frac{\kappa^n(x,\varphi)}{n}~\sum\limits_{m=1}^{\infty} \frac{\kappa^m(y,\varphi')}{m} >.
\end{align}
As $\kappa$ is small, we can ignore the higher order terms to get 
\begin{align}
<K(x,\varphi) K(y,\varphi')> \approx <\kappa(x,\varphi) \kappa(y,\varphi') > = \frac{1}{n_{\kappa}} \frac{1}{y} \delta(x-y) \delta(\varphi - \varphi').
\end{align}
Thus we have
\begin{align}
\sigma_{M,K}^2 = \frac{2\pi}{n_{\kappa}} \int\limits_{R}^{\infty}~dx~x^{2n+1} w(x)^2  - \frac{\pi^2}{n_{\kappa}^2 a^2} \Bigg(\int\limits_{R}^{\infty}~dx~x^{n+1} w(x) \Bigg)^2,\\
\sigma_{M,K} = \Bigg( \frac{2\pi}{n_{\kappa}} \int\limits_{R}^{\infty}~dx~x^{2n+1} w(x)^2  - \frac{\pi^2}{n_{\kappa}^2 a^2} \Bigg(\int\limits_{R}^{\infty}~dx~x^{n+1} w(x) \Bigg)^2 \Bigg)^{\frac{1}{2}}.
\end{align}
The variance is typically well behaved. However, as the first term depends on the source density and the second on the number of sources over which we average to obtain the convergence, $n_{\kappa} a$, it is theoretically possible to construct unreasonable combinations. For example, obtaining the convergence from only 1 source per bin while having a high source density would lead to an unreasonable result, as the contribution from the poorly constrained expectation value would dominate the other uncertainties. Naturally, such a combination would be avoided in real applications.
\\
\\
As we treat the case of no lensing, $<\kappa> = 0$ and we can thus average over the whole area $A$ to obtain our convergence estimate. Thus we have
\begin{align}
\sigma_{M,K} = \Bigg( \frac{2\pi}{n_{\kappa}} \int\limits_{R}^{\infty}~dx~x^{2n+1} w(x)^2  - \frac{\pi^2}{n_{\kappa}^2 A^2} \Bigg(\int\limits_{R}^{\infty}~dx~x^{n+1} w(x) \Bigg)^2 \Bigg)^{\frac{1}{2}}.
\end{align}
\\
\\
We have $< M_g^{(n)} > = 0$ and
\begin{align}
\sigma_{M,g}^2 = <|M_{g}^{(n)}|^2>\\
= \int\limits_{R}^{\infty}\int\limits_{R}^{\infty}~dx~dy~x W_R(x) y W_R(y) \int\limits_{0}^{2\pi}\int\limits_{0}^{2\pi}~d\varphi~d\varphi' \\ \nonumber 
(<g_t(x,\varphi) F_t(x,\varphi) g_t(y,\varphi') F_t(y,\varphi')> + <g_t(x,\varphi) F_t(x,\varphi) g_r(y,\varphi') F_r(y,\varphi')> \\ \nonumber 
+ <g_r(x,\varphi) F_r(x,\varphi) g_t(y,\varphi') F_t(y,\varphi')> + <g_r(x,\varphi) F_r(x,\varphi) g_r(y,\varphi') F_r(y,\varphi')>) \\ \nonumber
+ \int\limits_{R}^{\infty}\int\limits_{R}^{\infty}~dx~dy~x W_R(x) (2 W_R(y) -  y^{n+1} w(y)) \int\limits_{0}^{2\pi}\int\limits_{0}^{2\pi}~d\varphi~d\varphi'\\\nonumber
(<g_t(x,\varphi) F_t(x,\varphi) g_t(y,\varphi')> + <g_r(x,\varphi) F_r(x,\varphi) g_t(y,\varphi')>)\\\nonumber
+ \int\limits_{R}^{\infty}\int\limits_{R}^{\infty}~dx~dy~y W_R(y) (2 W_R(x) -  x^{n+1} w(x)) \int\limits_{0}^{2\pi}\int\limits_{0}^{2\pi}~d\varphi~d\varphi'\\\nonumber
(<g_t(y,\varphi') F_t(y,\varphi') g_t(x,\varphi)> + <g_r(y,\varphi') F_r(y,\varphi') g_t(x,\varphi)>)\\\nonumber
+ \int\limits_{R}^{\infty}\int\limits_{R}^{\infty}~dx~dy~(2 W_R(x) - x^{n+1} w(x)) (2 W_R(y) - y^{n+1} w(y)) \int\limits_{0}^{2\pi}\int\limits_{0}^{2\pi}~d\varphi~d\varphi'~<g_t(x,\varphi) g_t(y,\varphi')>.
\end{align}
In the absence of lensing, we expect the flexion and the shear to be uncorrelated. Therefore we have, using the relation shown in \citet{Goodman1960},
\begin{align}
<g_t(x,\varphi) F_t(x,\varphi) g_t(y,\varphi') F_t(y,\varphi')>\\
= (<g_t(x,\varphi) g_t(x,\varphi)> <F_t(x,\varphi) F_t(x,\varphi)> + <g_t(x,\varphi) g_t(x,\varphi)> < F_t(x,\varphi) >^2 \\
+ <F_t(x,\varphi) F_t(x,\varphi)> < g_t(x,\varphi) >^2) \frac{1}{n_{\gamma,\mathcal{F}}} \frac{1}{y} \delta(x - y) \delta(\varphi - \varphi')\\
= \frac{\sigma_{\gamma}^2 \sigma_{\mathcal{F}}^2}{n_{\gamma,\mathcal{F}}} \frac{1}{y} \delta(x - y) \delta(\varphi - \varphi'),
\end{align}
where $n_{\gamma,\mathcal{F}}$ is the number density of sources for which we have both shear and flexion information. As flexion is typically much harder to measure than shear, we can make the approximation $n_{\gamma,\mathcal{F}} \approx n_{\mathcal{F}}$. Using this approximation and making a similar calculation in the other cases, we have
\begin{align}
<g_t(x,\varphi) F_t(x,\varphi) g_t(y,\varphi') F_t(y,\varphi')> = \frac{\sigma_{\gamma}^2 \sigma_{\mathcal{F}}^2}{n_{\mathcal{F}}} \frac{1}{y} \delta(x - y) \delta(\varphi - \varphi'),\\
<g_r(x,\varphi) F_r(x,\varphi) g_r(y,\varphi') F_r(y,\varphi')> = \frac{\sigma_{\gamma}^2 \sigma_{\mathcal{F}}^2}{n_{\mathcal{F}}} \frac{1}{y} \delta(x - y) \delta(\varphi - \varphi'),\\
<g_t(x,\varphi) F_t(x,\varphi) g_r(y,\varphi') F_r(y,\varphi')> = <g_t(x,\varphi)> <g_r(y,\varphi')> <F_t(x,\varphi)> <F_r(y,\varphi')> = 0,\\
<g_t(x,\varphi) F_t(x,\varphi) g_t(y,\varphi')> = <g_t(x,\varphi) g_t(y,\varphi')> <F_t(x,\varphi)> = 0,\\
<g_r(x,\varphi) F_r(x,\varphi) g_t(y,\varphi')> = <g_r(x,\varphi) g_t(y,\varphi')> <F_r(x,\varphi)> = 0,
\end{align}
and thus
\begin{align}
\sigma_{M,g}^2 = \int\limits_{R}^{\infty}~dx~x W_R(x)^2 \int\limits_{0}^{2\pi}~d\varphi~2 \frac{\sigma_{\gamma}^2 \sigma_{\mathcal{F}}^2}{n_{\mathcal{F}}} + \int\limits_{R}^{\infty}~dx~\frac{1}{x} (2 W_R(x) - x^{n+1} w(x))^2 \int\limits_{0}^{2\pi}~d\varphi~\frac{\sigma_{\gamma}^2}{n_{\gamma}} 
\end{align}
and we have the standard deviation
\begin{align}
\sigma_{M,g} = \Bigg( \frac{4\pi}{n_{\mathcal{F}}} \sigma_{\mathcal{F}}^2 \sigma_{\gamma}^2 (\int\limits_{R}^{\infty}~dx~x W_R(x)^2) + \frac{2\pi}{n_{\gamma}} \sigma_{\gamma}^2 \int\limits_{R}^{\infty}~dx~\frac{1}{x} (2 W_R(x) - x^{n+1} w(x))^2  \Bigg)^{\frac{1}{2}}.
\end{align}
\\
\\
We have $<M_F^{(n)}> = 0$ and 
\begin{align}
\sigma_{M,F}^2 = <|M_{F}^{(n)}|^2>\\
= \int\limits_{R}^{\infty} \int\limits_{R}^{\infty}~dx~dy~x W_R(x) y W_R(y) \int\limits_{0}^{2\pi} \int\limits_{0}^{2\pi}~d\varphi~d\varphi'~ <F_t(x,\varphi) F_t(y,\varphi') >
\end{align}
and using
\begin{align}
<F_t(x,\varphi) F_t(y,\varphi') > = \frac{\sigma_{\mathcal{F}}^2}{n_{\mathcal{F}}} \frac{1}{y} \delta(x - y) \delta(\varphi - \varphi')
\end{align}
we have
\begin{align}
\sigma_{M,F}^2 = \frac{2\pi}{n_{\mathcal{F}}} \sigma_{\mathcal{F}}^2 \int\limits_{R}^{\infty}~dx~x W_R(x)^2 
\end{align}
and the standard deviation
\begin{align}
\sigma_{M,F} = \sqrt{\frac{2\pi}{n_{\mathcal{F}}}} \sigma_{\mathcal{F}} \Bigg( \int\limits_{R}^{\infty}~dx~x W_R(x)^2 \Bigg)^{\frac{1}{2}}.
\end{align}
\\
\\
We have $< M_{F,G}^{(n)} > = 0$ and
\begin{align}
\sigma_{M,F,G}^2 = <|M_{F,G}^{(n)}|^2>\\
= \int\limits_{R}^{\infty} \int\limits_{R}^{\infty}~dx~dy~x W_R(x) y W_R(y) \int\limits_{0}^{2\pi} \int\limits_{0}^{2\pi}~d\varphi~d\varphi'~( < g_r(x,\varphi) F_r(x,\varphi) g_r(y,\varphi') F_r(y,\varphi') > \\ \nonumber
+ \frac{1}{2} < g_r(x,\varphi) F_r(x,\varphi) F_t(y,\varphi') > + \frac{1}{2} < g_r(x,\varphi) F_r(x,\varphi) G_t(y,\varphi') > + \frac{1}{2} < F_t(x,\varphi) g_r(y,\varphi') F_r(y,\varphi') > \\ \nonumber  
+ \frac{1}{2} < G_t(x,\varphi) g_r(y,\varphi') F_r(y,\varphi') > + \frac{1}{4} < F_t(x,\varphi) F_t(y,\varphi') > + \frac{1}{4} < F_t(x,\varphi) G_t(y,\varphi') > \\ \nonumber
 + \frac{1}{4} < G_t(x,\varphi) F_t(y,\varphi') > + \frac{1}{4} < G_t(x,\varphi) G_t(y,\varphi') > ) \\ \nonumber
- \int\limits_{R}^{\infty} \int\limits_{R}^{\infty}~dx~dy~x W_R(x) V_R(y) \int\limits_{0}^{2\pi} \int\limits_{0}^{2\pi}~d\varphi~d\varphi'~(< g_r(x,\varphi) F_r(x,\varphi) F_t(y,\varphi') > \\ \nonumber
+ < g_r(x,\varphi) F_r(x,\varphi) G_t(y,\varphi') > - 2 < g_r(x,\varphi) F_r(x,\varphi) g_t(y,\varphi') F_t(y,\varphi') > \\\nonumber
+ \frac{1}{2} < F_t(x,\varphi) F_t(y,\varphi') > + \frac{1}{2} < F_t(x,\varphi) G_t(y,\varphi') > - < F_t(x,\varphi) g_t(y,\varphi') F_t(y,\varphi') > \\\nonumber
+ \frac{1}{2} < G_t(x,\varphi) F_t(y,\varphi') > + \frac{1}{2} < G_t(x,\varphi) G_t(y,\varphi') > - < G_t(x,\varphi) g_t(y,\varphi') F_t(y,\varphi') >) \\\nonumber
- \int\limits_{R}^{\infty} \int\limits_{R}^{\infty}~dx~dy~V_R(x) y W_R(y) \int\limits_{0}^{2\pi} \int\limits_{0}^{2\pi}~d\varphi~d\varphi'~(< g_r(y,\varphi') F_r(y,\varphi') F_t(x,\varphi) > \\ \nonumber
+ < g_r(y,\varphi') F_r(y,\varphi') G_t(x,\varphi) > - 2 < g_r(y,\varphi') F_r(y,\varphi') g_t(x,\varphi) F_t(x,\varphi) > \\\nonumber
+ \frac{1}{2} < F_t(y,\varphi') F_t(x,\varphi) > + \frac{1}{2} < F_t(y,\varphi') G_t(x,\varphi) > - < F_t(y,\varphi') g_t(x,\varphi) F_t(x,\varphi) > \\\nonumber
+ \frac{1}{2} < G_t(y,\varphi') F_t(x,\varphi) > + \frac{1}{2} < G_t(y,\varphi') G_t(x,\varphi) > - < G_t(y,\varphi') g_t(x,\varphi) F_t(x,\varphi) >) \\\nonumber
+ \int\limits_{R}^{\infty} \int\limits_{R}^{\infty}~dx~dy~V_R(x) V_R(y) \int\limits_{0}^{2\pi} \int\limits_{0}^{2\pi}~d\varphi~d\varphi'~(< F_t(x,\varphi) F_t(y,\varphi') >
+ <F_t(x,\varphi) G_t(y,\varphi')> \\\nonumber
- 2 <F_t(x,\varphi) g_t(y,\varphi') F_t(y,\varphi')> + <G_t(x,\varphi) F_t(y,\varphi')> + <G_t(x,\varphi) G_t(y,\varphi')> \\\nonumber
-2 <G_t(x,\varphi) g_t(y,\varphi') F_t(y,\varphi')>
- 2 < g_t(x,\varphi) F_t(x,\varphi) F_t(y,\varphi') > \\\nonumber
- 2 < g_t(x,\varphi) F_t(x,\varphi) G_t(y,\varphi') > + 4 < g_t(x,\varphi) F_t(x,\varphi) g_t(y,\varphi') F_t(y,\varphi') > ).
\end{align}
We again assume that the first and second flexions and the shear are uncorrelated in the absence of lensing, thus we can use the previously derived relations and we also have
\begin{align}
<G_t(x,\varphi) G_t(y,\varphi') > = \frac{\sigma_{\mathcal{G}}^2}{n_{\mathcal{G}}} \frac{1}{y} \delta(x-y) \delta(\varphi - \varphi'),\\
< g_r(x,\varphi) F_r(x,\varphi) F_t(y,\varphi') > = < g_r(x,\varphi) F_r(x,\varphi) G_t(y,\varphi') > = < F_t(x,\varphi) G_t(y,\varphi') > =  0,\\
< F_t(x,\varphi) g_t(y,\varphi') F_t(y,\varphi') > = <g_t(y,\varphi')> <F_t(x,\varphi) F_t(y,\varphi')> = 0,\\
< G_t(x,\varphi) g_t(y,\varphi') F_t(y,\varphi') > = 0.
\end{align}
This gives us
\begin{align}
\sigma_{M,F,G}^2 = 2\pi (\frac{\sigma_{\gamma}^2 \sigma_{\mathcal{F}}^2}{n_{\mathcal{F}}} + \frac{\sigma_{\mathcal{F}}^2}{4 n_{\mathcal{F}}} + \frac{\sigma_{\mathcal{G}}^2}{4 n_{\mathcal{G}}}) \int\limits_{R}^{\infty}~dx~x W_R(x)^2 - 2\pi (\frac{\sigma_{\mathcal{F}}^2}{n_{\mathcal{F}}} + \frac{\sigma_{\mathcal{G}}^2}{n_{\mathcal{G}}}) \int\limits_{R}^{\infty}~dx~W_R(x) V_R(x) \nonumber \\
+ 2\pi (\frac{\sigma_{\mathcal{F}}^2}{n_{\mathcal{F}}} + \frac{\sigma_{\mathcal{G}}^2}{n_{\mathcal{G}}} + 4\frac{\sigma_{\gamma}^2 \sigma_{\mathcal{F}}^2}{n_{\mathcal{F}}}) \int\limits_{R}^{\infty}~dx~\frac{1}{x} V_R(x)^2,\\
\sigma_{M,F,G} = \sqrt{2\pi} \Bigg( (\frac{\sigma_{\gamma}^2 \sigma_{\mathcal{F}}^2}{n_{\mathcal{F}}} + \frac{\sigma_{\mathcal{F}}^2}{4 n_{\mathcal{F}}} + \frac{\sigma_{\mathcal{G}}^2}{4 n_{\mathcal{G}}}) \int\limits_{R}^{\infty}~dx~x W_R(x)^2 - (\frac{\sigma_{\mathcal{F}}^2}{n_{\mathcal{F}}} + \frac{\sigma_{\mathcal{G}}^2}{n_{\mathcal{G}}}) \int\limits_{R}^{\infty}~dx~W_R(x) V_R(x) \nonumber \\
+ (\frac{\sigma_{\mathcal{F}}^2}{n_{\mathcal{F}}} + \frac{\sigma_{\mathcal{G}}^2}{n_{\mathcal{G}}} + 4\frac{\sigma_{\gamma}^2 \sigma_{\mathcal{F}}^2}{n_{\mathcal{F}}}) \int\limits_{R}^{\infty}~dx~\frac{1}{x} V_R(x)^2   \Bigg)^{\frac{1}{2}}.
\end{align}

\section{Convergence, shear, and flexion for NFW profiles}\label{Appendix: NFW formulas}

Navarro, Frenk, and White showed that spherically averaged cold dark matter (CDM) halo density profiles can be fitted over two decades in radius by the NFW density profile \citep{Navarro1996,Navarro1997}
\begin{eqnarray}
\frac{\rho(r)}{\rho_{\rm crit}} \,=\, \frac{\delta_{c}}{(r/r_{s})\, (1 + r/r_{s})^{2}},
\end{eqnarray}
where $r_{s}$ is a scale radius, $\delta_{c}$ is a characteristic dimensionless density, and $\rho_{\rm crit} \,=\, 3 H^{2}/8 \pi G$ is the critical density for closure. This profile leads to the convergence formula \citep{Bartelmann1996,Wright2000}
\begin{eqnarray}
\kappa_{\rm NFW}(y) &\,=\,& \frac{2 \rho_{\rm crit}\, \delta_{c}\, r_{s}}{\Sigma_{\rm crit}}\, \frac{f(y)}{y^{2} - 1},
\end{eqnarray}
where $y \,=\, \xi/r_{s}$ and
\begin{eqnarray}
f(y) \,=\,\left\{ 
\begin{array}{ll}
1 - \frac{2}{\sqrt{1 - y^{2}}} \arctanh \Big(\sqrt{\frac{1 - y}{1 + y}}\, \Big),\,\, & y < 1 \\
\frac{y^2 - 1}{3},\,\, & y = 1 \\
1 - \frac{2}{\sqrt{y^{2} - 1}} \arctan \Big(\sqrt{\frac{y - 1}{y + 1}}\, \Big),\,\, & y > 1.
\end{array} \right.
\end{eqnarray}
The NFW profile is spherically symmetric and therefore $\gamma_{r, \rm NFW} = 0$ \citep[see e.g.][]{Schneider2006}. The tangential shear is \citep{Wright2000}
\begin{eqnarray}
\gamma_{t, \rm NFW} \,=\, \frac{\rho_{\rm crit}\, \delta_{c}\, r_{s}}{\Sigma_{\rm crit}}\, g(y),
\end{eqnarray}
where
\begin{eqnarray}
g(y) \,=\, \left\{
\begin{array}{ll}
\frac{8}{y^{2} \sqrt{1 - y^{2}}}\, \arctanh \Big(\sqrt{\frac{1 - y}{1 + y}}\, \Big) + \frac{4}{y^{2}} \log \Big(\frac{y}{2}\Big) 
- \frac{2}{y^{2} - 1} + \frac{4}{(y^{2} - 1)\, \sqrt{1 - y^{2}}}\, \arctanh \Big(\sqrt{\frac{1 - y}{1 + y}}\, \Big),\,\, & y < 1 \\ & \\
\frac{10}{3} + 4 \log \Big(\frac{1}{2}\Big),\,\, & y = 1 \\ & \\
\frac{8}{y^{2} \sqrt{y^{2} - 1}}\, \arctan \Big(\sqrt{\frac{y - 1}{1 + y}}\, \Big) + \frac{4}{y^{2}} \log \Big(\frac{y}{2}\Big) 
- \frac{2}{y^{2} - 1} + \frac{4}{(y^{2} - 1)^{3/2}}\, \arctan \Big(\sqrt{\frac{y - 1}{1 + y}}\, \Big),\,\, & y > 1.
\end{array} \right.
\end{eqnarray}
We can use the expressions for first and second flexion for a NFW profile derived in \citet{Bacon2006} to obtain
\begin{eqnarray}
\mathcal{F}_{t, \rm NFW} = \frac{2 \rho_{\rm crit}\, \delta_{c} D_{\rm ol}}{\Sigma_{\rm crit}(y^{2} - 1)^{2}}\, \Big[2 y\,f(y) - h(y) \Big], \\
\mathcal{G}_{t, \rm NFW} = - \frac{2 \rho_{\rm crit}\, \delta_{c} D_{\rm ol}}{\Sigma_{\rm crit}}\, 
\Big[ \frac{8}{y^{3}}\, \log\Big(\frac{y}{2}\Big) + \frac{(3 (1 - 2 y^{2})/y) + l(y)}{(y^{2} - 1)^{2}} \Big],
\end{eqnarray}
where
\begin{eqnarray}
h(y) &\,=\,& \left\{
\begin{array}{ll}
\frac{2 y}{\sqrt{1 - y^{2}}}\, \arctanh \Big(\sqrt{\frac{1 - y}{1 + y}}\,\Big) - \frac{1}{y},\,\, & y < 1 \\
\frac{2 y (y^2 -1)}{3} - 0.4~(y^2 - 1)^2,\,\, & y = 1 \\
\frac{2 y}{\sqrt{y^{2} - 1}}\, \arctan \Big(\sqrt{\frac{y - 1}{y + 1}}\,\Big) - \frac{1}{y},\,\, & y > 1,
\end{array} \right. \\
l(y) &\,=\,& \left\{
\begin{array}{ll}
\Big(\frac{8}{y^{3}} - \frac{20}{y} + 15 y\Big)\, \frac{2}{\sqrt{1 - y^{2}}}\, \arctanh \Big(\sqrt{\frac{1 - y}{1 + y}}\,\Big),\,\, & y < 1 \\
\frac{1}{y}\Big(\frac{94}{15} y (y^2 - 1)^2 - 3(1 - 2 y^2)\Big),\,\, & y = 1 \\  
\Big(\frac{8}{y^{3}} - \frac{20}{y} + 15 y\Big)\, \frac{2}{\sqrt{y^{2} - 1}}\, \arctan \Big(\sqrt{\frac{y - 1}{y + 1}}\,\Big),\,\, & y > 1.
\end{array} \right.
\end{eqnarray}
The radial flexion components are both zero.

\bsp
\label{lastpage}

\end{document}